\documentclass[12pt]{iopart}

\usepackage[margin=1in]{geometry}
\usepackage{graphicx}
\usepackage{bbm}
\usepackage{bm}
\usepackage{enumitem}
\usepackage{booktabs}
\usepackage{microtype}
\usepackage{xcolor}
\usepackage{cite}

\makeatletter
\expandafter\let\csname equation*\endcsname\relax
\expandafter\let\csname endequation*\endcsname\relax
\makeatother

\usepackage{amsmath,amssymb,amsfonts,amsthm,mathtools}
\usepackage[nameinlink,capitalize]{cleveref}

\makeatletter
\long\def\@makecaption#1#2{%
  \vskip\abovecaptionskip
  \begin{center}%
  \begin{minipage}{0.94\linewidth}%
  \small #1. #2%
  \end{minipage}%
  \end{center}%
  \vskip\belowcaptionskip}
\makeatother

\usepackage{physics}
\usepackage{mathtools}

\numberwithin{equation}{section}

\newtheorem{theorem}{Theorem}[section]

\newtheorem{definition}[theorem]{Definition}

\newcommand{\ii}{\mathrm i}
\newcommand{\ApplyQ}{\operatorname{ApplyQ}}
\newcommand{\Applyq}{\operatorname{Applyq}}
\newcommand{\ApplyA}{\operatorname{ApplyA}}
\newcommand{\Hcal}{\mathcal H}
\newcommand{\Acal}{\mathcal A}
\newcommand{\Gcal}{\mathcal G}

\begin{document}

\title[]{A matrix free action of the Ashtekar-Lewandowski volume operator of loop quantum gravity}

\author{Waleed Sherif}

\address{Institute for Quantum Gravity, Department of Physics, Friedrich-Alexander-Universit\"{a}t Erlangen-N\"{u}rnberg (FAU), Staudtstraße 7, 91058 Erlangen, Germany}

\ead{waleed.sherif@fau.de}

%
%

\begin{abstract}
The Ashtekar–Lewandowski (AL) volume operator of loop quantum gravity is central to the Hamiltonian constraint, but its vertex action is usually obtained from dense spectral decompositions of finite recoupling matrices, obstructing numerical analysis on large kinematical Hilbert spaces or high-valence vertices. We formulate a matrix free action of the $\mathrm{SU}(2)$ AL vertex volume operator in standard recoupling basis, making use of the Brunnemann-Thiemann expression for the oriented AL volume density $Q_v$, whose matrix elements can be generated locally from recoupling theory without forming the full matrix. Based on the Balakrishnan–Stieltjes representation of $(Q_v^2)^{1/4}$, we approximate the volume by shifted-resolvent quadrature (SRQ). The resulting action uses only repeated applications of $Q_v$ and shifted positive linear solves, making it compatible with multi-shift Krylov methods. We prove exact preservation of the volume kernel, provide operator-norm and residual error estimates, discuss sector-wise scaling bounds, and validate the method on an embedded $K_5$ graph at small spin cutoffs against exact dense local-block operators. Numerical simulations show rapid convergence of vertex expectation values, controlled dependence on bound parameters, and exact preservation of zero-volume modes. We further demonstrate matrix free Monte Carlo estimates at doubled-spin cutoff $2j=250000$, beyond dense materialisation, and show that SRQ can be combined with stochastic Lanczos quadrature to estimate fixed-sector volume spectral measures without dense volume matrices.
\end{abstract}

%
%

\section{\label{sec:introduction}Introduction}
Loop quantum gravity (LQG) is a background independent approach to quantising general relativity whereby quantum states of spatial geometry are described by spin network functions (SNFs) on kinematical Hilbert spaces defined on embedded spatial graphs~\cite{Rovelli:1997yv,Thiemann:2001gmi,Thiemann:2007pyv,Ashtekar:2004eh,Ashtekar:2021kfp}. At the kinematical level, geometric operators such as area and volume operators are shown to have discrete spectra~\cite{Rovelli:1994ge,Ashtekar:1996eg,Ashtekar:1997fb}. In the canonical approach to LQG, physical states are obtained by simultaneously finding the kernel of three quantum constraints, namely the Gauß constraint $\hat{G}$, the spatial diffeomorphism constraint $\hat{D}$ and the quantum Hamilton constraint $\hat{H}$~\cite{Thiemann:1996aw,Thiemann:1996av,Thiemann:1996ay,Thiemann:2007pyv}.

Significant effort, from both analytical and numerical aspects, has been devoted to finding physical states, in particular near-kernel states of the quantum Hamilton constraint~\cite{Thiemann:1996aw,Thiemann:1996av,Thiemann:1996ay,Thiemann:2007pyv}. A central non-polynomial operator which appears in the construction of $\hat{H}$ is the volume operator. In this work, we consider the Ashtekar-Lewandowski regularization of the volume operator whereby it is obtained by a background-independent regularization of the classical volume functional and is defined as a sum of positive vertex operators on SNFs~\cite{Ashtekar:1997fb,Thiemann:1996at}. At each vertex, the Ashtekar-Lewandowski operator is the square root of the absolute value of a single oriented volume-density operator,
\begin{equation}
        \hat V^{\mathrm{AL}}_v
        \propto
        \sqrt{\left|
        \sum_{I,J,K}\epsilon_{IJK}(v)\,
        \epsilon_{ijk}X_I^iX_J^jX_K^k
        \right|}.
\end{equation}
Other regularizations of the volume operator differ in their expression. For example, in the such as the the Rovelli-Smolin-De Pietri regularization, the absolute value is taken at the level of the individual triple graspings before the outer square root~\cite{Rovelli:1994ge,DePietri:1996tvo,Ashtekar:1997fb}, changing its sensitivity to sign factors, and it further does not depend on the graph embedding.

Recent numerical developments have approached the dynamics of canonical LQG from different complementary directions. To name a few, one line of work has developed tools for the graph-changing action of Thiemann's Hamiltonian constraint, including the action of the volume operator, and has used them to study graph-changing dynamics and solutions of the constraint~\cite{Guedes:2025taming,Guedes:2025graphchanging}. A second line of work has applied scalable variational methods to graph-preserving Abelianised LQG and quantum reduced loop gravity, where near-kernel states can be searched for in reduced or truncated settings~\cite{Sahlmann:2024pba,Sahlmann:2024kat,Sahlmann:2026qvs,Makinen:2026wwp}. Although these approaches differ in scope, they expose a common numerical bottleneck. Large scale numerical analysis requires repeated and efficient evaluation of the relevant quantum constraints and geometric operators, while the construction and evaluation of these operators becomes increasingly expensive as one moves toward the full $\mathrm{SU}(2)$ theory and beyond simplistic models.

For the Hamiltonian constraint $\hat{H}$, this difficulty is tied in particular to the volume operator, which plays a central role in the definition of the constraint. Its action on a given spin-network function, however, remains highly non-trivial and difficult to evaluate numerically~\cite{Thiemann:1996aw,Thiemann:1996av,Thiemann:1996ay,Brunnemann:2004xi,brunnemann-and-rideout}. It therefore represents a limiting component both for graph-changing computations of the Hamiltonian and for any future extension of scalable variational methods to the $\mathrm{SU}(2)$ setting.

Earlier semiclassical strategies for avoiding direct spectral evaluation in more restricted settings, including complexifier coherent state calculations of volume-dependent expectation values in matter Hamiltonians coupled to quantum geometry and algebraic quantum gravity and LQG semiclassical perturbation theory for matrix elements of fractional powers of the volume operator, have been developed ~\cite{Sahlmann:2002qj,Giesel:2006uk}. However, despite such substantial efforts, these do not provide a general matrix free and scalable action of $V_v$ on arbitrary spin-network data. 

Closed form spin network matrix elements of the volume operator, derived by Thiemann in \cite{Thiemann:1996at}, and recoupling theoretic simplifications by Brunnemann and Thiemann \cite{Brunnemann:2004xi} made large parts of the spectral analysis computationally accessible by replacing prohibitively large sums of general $6j$-symbols with compact formulas based on the Elliot-Biedenharn identity. Higher valent numerical studies \cite{brunnemann-and-rideout,BrunnemannRideout:2007xk} demonstrated both the importance of the vertex sign configuration and the difficulty of the spectral problem beyond the 4-valent case.

Such efforts, however, mainly concentrated on the \emph{volume density operator}, whereby the vertex volume operator, required for scalable variational methods, remains largely untreated. Recently, however, a complementary integral transform approach in a Fock representation setting, in which non-polynomial metric density functions, including volume functionals, are defined as quadratic forms using Weyl quantisation based on Fourier transforms has been developed~\cite{ThiemannFock2024}. Although this lies in a different representation-theoretic setting from the finite-dimensional Ashtekar-Lewandowski recoupling blocks studied here, it is nevertheless directly applicable, and thus demonstrates that integral-transform representations of volume-type functions can be obtained in a manner beyond explicit spectral decomposition.

In this work, following from a similar integral transform approach principle as in~\cite{ThiemannFock2024}, we draw from the advances in the use of neural quantum states in variational Monte Carlo methods~\cite{Sorella:1998,Sorella:2001,Carleo:2016svm} as well as modern methods in automatic differentiation~\cite{Baydin:2018,jax2018github}, in particular stochastic-reconfiguration and matrix free Jacobian-vector and vector-Jacobian products (JVPs and VJPs respectively)~\cite{Amari:1998,Sorella:2001,Baydin:2018,jax2018github}, to construct a matrix free algorithm for applying the vertex volume operator on SNFs on graphs with vertices of arbitrary valence and arbitrary intertwiner space dimensions. This is done by viewing the developed numerical methods for the volume density operator as an action callback, which can be applied repeatedly, and the vertex volume operator is then applied through shifted linear solves rather than through explicit spectral decomposition.

%
%

\section{\label{sec:volumeoperatorreview}The Ashtekar-Lewandowski volume operator}
To fix the notation and terminology, we begin with a brief review of the Ashtekar-Lewandowski volume operator in LQG~\cite{Ashtekar:1997fb,Thiemann:1996at}. In what follows, we denote by $\gamma$ and embedded oriented spatial graph and we denote by $v \in V(\gamma)$ a vertex in $\gamma$ of valence $N$. All incident edges are treated as outgoing, and incoming edges are converted into outgoing ones. Denote the incident spins at a given $N$-valent vertex by $\bm{j} = (j_1, \cdots, j_N)$ with $j_I \in \tfrac{1}{2}\mathbb{N}_0$, with representation space $V_{j_I}$. The finite-dimensional gauge invariant intertwiner space at $v$ is denoted as
\begin{equation}
    \Hcal_v(\bm{j}) = \mathrm{Inv}_{\mathrm{SU}(2)} \left(\bigotimes_{I=1}^N V_{j_I}\right),
\end{equation}
which we denote its dimension by $d_v(\bm{j}) = \dim\Hcal_v(\bm{j})$. Fix a recoupling tree hereby denoted by $T$. For definiteness, one may take the left associated tree, although none of the constructions that follow depend on this choice. A recoupling basis vector is denoted by $\ket{\bm{a}} = \ket{a_2, a_3, \cdots, a_{N-1}}_T$, where intermediate spins satisfy the usual Clebsch-Gordan conditions. For the left associated tree, set $a_1 = j_1$, then $\abs{a_{r-1} - j_r} \leq a_r \leq a_{r_1} + j_r$, with $a_{r-1} + j_r + a_r \in \mathbb{Z}$, for the relevant range of $r$.

Denote by $\Acal_{\bm{j}, T}$ the finite set of admissible recoupling labels, and we will assume the standard spin network scalar product in which the chosen normalised recoupling states are orthonormal~\cite{DePietri:1996tvo,Thiemann:2007pyv}. Gauge invariance is implemented by the final coupling to total spin zero. Explicitly, if one denotes by $X_I^i, i=1, 2, 3$ the self-adjoint $su(2)$ angular momentum generators acting on the $I$\textsuperscript{th} tensor factor and trivially on others, the Gauß constraint on $\Hcal_v(\bm{j})$ is simply $\sum_{I=1}^N \bm{X}_I = 0$. For distinct labels $I, J, K$, define
\begin{equation}
    G_{IJK} := \epsilon_{ijk}X_I^i X_J^j X_K^k.
\end{equation}
Since the generators acting on distinct tensor factors commute, $G_{IJK}$ is self-adjoint and is totally antisymmetric in the edge labels. Now, let $\epsilon_{IJK}(v) = \mathrm{sgn}\,\det(\dot{e}_I(v), \dot{e}_J(v), \dot{e}_K)$, with $\epsilon_{IJK}(v) \in \{0, \pm 1\}$ be the Ashtekar-Lewandowski orientation sign of the ordered edge triple at $v$~\cite{Ashtekar:1997fb,Thiemann:1996at}. In the unordered-triple convention, define the dimensionless Ashtekar-Lewandowski oriented density, on the gauge invariant subspace, as
\begin{equation}
    Q_v := \sum_{1 \leq I < J < K < N} \sigma_{IJK}(v) G_{IJK},
\end{equation}
where $\sigma_{IJK}(v) = \epsilon_{IJK} - \epsilon_{IJN} + \epsilon_{IKN} - \epsilon_{JKN}$. Throughout the rest of the paper, \(Q_v\) denotes this Ashtekar-Lewandowski oriented volume density. With the usual ordered-triple normalisation, the corresponding dimensional operator is
\begin{equation}
    \hat{q}_v^{\mathrm{AL}} = \frac{(8\pi\gamma\ell_P^2)^3}{48} \sum_{I,J,K=1}^N \epsilon_{IJK}(v) \epsilon_{ijk} X_I^i X_J^j X_K^k = \frac{(8\pi\gamma\ell_P^2)^3}{48} Q_v,
\end{equation}
where the last equality uses antisymmetry and passes to unordered triples. The vertex volume then takes the form \cite{Ashtekar:1997fb}
\begin{equation}
    \hat{V}^{\mathrm{AL}}_v = \sqrt{\abs{\hat{q}_v^{\mathrm{AL}}}} = C_V \sqrt{\abs{Q_v}}, \quad C_V := \left(\frac{(8\pi\gamma\ell_P^2)^3}{48}\right)^{\tfrac{1}{2}}.
\end{equation}
The equation above is the Ashtekar-Lewandowski vertex volume considered in this paper\footnote{It is important to note that different expressions for the volume operator exists, for example the Rovelli-Smolin-De Pietri regularisation, which places the absolute value on the terms inside of the sum in $Q_v$ rather than after the summation and further does not depend on the embedding of the graph \cite{Rovelli:1994ge,DePietri:1996tvo}.}. Differences of normalization within this Ashtekar-Lewandowski convention are absorbed into \(C_V\). Lastly, since \(Q_v\) is a finite-dimensional self-adjoint operator on \(\Hcal_v(\bm{j})\), its absolute value is defined by the spectral theorem and one has $\sqrt{\abs{Q_v}} = (Q_v^2)^{1/4}$. We therefore denote the positive operator $A_v := Q_v^2 \geq 0,$
such that $\hat V_v^{\mathrm{AL}} = C_V A_v^{1/4}$. Note that in many recoupling bases, the matrix of \(Q_v\) is purely imaginary and antisymmetric \cite{DePietri:1996tvo,Brunnemann:2004xi,brunnemann-and-rideout}, \(Q_v = \ii q_v\), with \(q_v\) real antisymmetric. In that case, $A_v = Q_v^2 = -q_v^2 = q_v^\top q_v \geq 0$, which is often preferable for real arithmetic implementations. In all cases, this does not affect the present work as one simply applies $A_v$ as $-q_v (q_v \psi)$ for some $\psi \in \Hcal_v(\bm{j})$. We note that, all occurrences of \(Q_v\), \(A_v=Q_v^2\), \(\bar A_v\), and \(\Lambda_v\) below refer to the Ashtekar-Lewandowski oriented-density construction unless explicitly stated otherwise.

\section{\label{sec:matrixfreevertexvolume}Matrix free vertex volume action}

Numerical simulations aiming to solve the quantum Hamilton constraint must confront the difficulty of applying the Ashtekar-Lewandowski vertex volume operator on a given basis element. In this section, we develop an algorithmic approach to applying this operator. The known recoupling formulas \cite{Thiemann:1996at,DePietri:1996tvo,Brunnemann:2004xi} give matrix elements
\begin{equation}
    Q_{\bm{ba}}^{(v,T)} = \langle \bm{b}, Q_v \,\bm{a} \rangle_T = \sum_{I < J < K < N} \sigma_{IJK} \langle \bm{b}, G_{IJK} \,\bm{a} \rangle_T.
\end{equation}
In this work, it is important to conceptually isolate their structural consequence. For each triple $(I, J, K)$ and each input channel $\bm{a}$, there is a finite set $\Gcal_{IJK}(\bm{a}) \subset \Acal_{\bm{j}, T}$ of output channels and computable amplitudes $g_{\bm{ba}}^{IJK}:= \langle \bm{b}, Q_v \, \bm{a}\rangle_T$, where $\bm{b} \in \Gcal_{IJK}(\bm{a})$, with $g_{\bm{ba}}^{IJK} = 0$ outside this set. This simplification of \cite{Brunnemann:2004xi} drastically reduces the cost of computing these amplitudes by eliminating general $6j$-symbol summations in favour of compact expressions. To formalise notation, we denote the matrix free primitive of such an operation as follows.

\begin{definition}[Recoupling-action oracle]\label{def:ApplyQ}
A recoupling-action oracle for the vertex $(v,\bm j,T)$ is a routine
\begin{equation}
  \ApplyQ_v : \mathbb{C}^{\Acal_{\bm j,T}} \to \mathbb{C}^{\Acal_{\bm j,T}},
\end{equation}
which, for an input coefficient vector $\psi=(\psi_{\bm a})$, returns the vector $\phi=Q_v\psi$ by the scatter-add rule
\begin{equation}\label{eq:ApplyQ-scatter}
  \phi_{\bm b} = \sum_{\bm a\in\Acal_{\bm j,T}}
   \sum_{I<J<K<N}\sigma_{IJK}\,g^{IJK}_{\bm b\bm a}\,\psi_{\bm a}.
\end{equation}
The routine is matrix free if the entries $Q_{\bm b\bm a}^{(v,T)}$ are generated only locally as needed and the full $d_v\times d_v$ matrix of $Q_v$ is never stored.
\end{definition}

The oracle may internally cache local recoupling coefficients, admissibility intervals, or reduced signs. It must not, however, materialise the dense matrix of $Q_v$ or any dense matrix derived from it. Now, given $\ApplyQ_v$, define
\begin{equation}
    \ApplyA_v(\psi) := \ApplyQ_v(\ApplyQ_v(\psi)),
\end{equation}
which in exact arithmetic is precisely $A_v \psi = Q_v^2 \psi$. If one uses the real antisymmetric convention $Q_v = \ii q_v$, the corresponding positive action is $\ApplyA_v(\psi) := -\Applyq_v(\Applyq_v(\psi))$.

Notation aside, we now formulate a matrix free vector-volume action such that, given \(\psi \in \Hcal_v(\bm{j})\), one computes or approximates $\hat V_v^{\mathrm{AL}}\psi = C_V A_v^{1/4}\psi$ using vector operations and invocations of \(\ApplyQ_v\), without constructing a dense matrix of \(Q_v\), \(A_v=Q_v^2\), \(A_v^{1/4}\), or any intermediaries. This analytically can be achieved using standard Balakrishnan-Stieltjes representation of fractional powers of positive operators, widely used in numerical methods for matrix functions and rational approximation \cite{Balakrishnan:1960,Higham:2008,HaleHighamTrefethen2008,TrefethenWeideman2014}. At the scalar level, for $0<\alpha<1$ and $\lambda\ge0$,
\begin{equation}\label{eq:scalar-stieltjes}
  \lambda^\alpha
  =\frac{\sin(\pi\alpha)}{\pi}
  \int_0^\infty t^{\alpha-1}\frac{\lambda}{t+\lambda}\, dt,
\end{equation}
with both sides understood to vanish at $\lambda=0$.  By the spectral theorem,
the corresponding finite-dimensional operator identity for a positive
self-adjoint operator $A\ge0$ is
\begin{equation}\label{eq:operator-stieltjes}
  A^\alpha
  =\frac{\sin(\pi\alpha)}{\pi}
  \int_0^\infty t^{\alpha-1}A(t\mathbb{I}+A)^{-1}\, dt,
\end{equation}
where the integral converges in operator norm.  After a change
of variables $t=e^s$, this becomes
\begin{equation}\label{eq:operator-log-stieltjes}
  A^\alpha
  =\frac{\sin(\pi\alpha)}{\pi}
  \int_{-\infty}^{\infty}e^{\alpha s}A(e^s\mathbb{I}+A)^{-1}\, ds.
\end{equation}
This representation is useful here because it replaces a fractional power by
a continuum of shifted resolvents.  Each resolvent involves the positive
operator $A$ only through products with vectors, and the factor $A$ appearing
inside the integrand makes the kernel behaviour explicit.

For the Ashtekar-Lewandowski vertex volume operator one applies \eqref{eq:operator-log-stieltjes}
with \(A=A_v=Q_v^2\) and \(\alpha=1/4\). Thus,
\begin{equation}
    \hat{V}^{\mathrm{AL}}_v \psi
    =
    C_V \frac{1}{\pi \sqrt{2}}
    \int_{-\infty}^\infty e^{s/4}
    A_v (e^s \mathbb{I} + A_v)^{-1} \psi \, ds.
\end{equation}
For notational brevity, when no confusion is possible we subsequently write \(\hat V_v\) for \(\hat V_v^{\mathrm{AL}}\). Since one can represent the action of $A_v$ in a matrix free manner, the equation above already gives a matrix free expression for the volume operator, given one can resolve the integral. In order to do this, let $\Lambda_v > 0$ satisfy $\norm{Q_v} \leq \Lambda_v$ and define the scaled operator
\begin{equation}
    \Bar{A}_v := \frac{A_v}{\Lambda_v^2} = \frac{Q_v^2}{\Lambda_v^2}.
\end{equation}
Then, $0 \leq \Bar{A}_v \leq 1$ and $A_v^{1/4} = \sqrt{\Lambda_v} \Bar{A}_v^{1/4}$. For $\alpha = 1/4$, then
\begin{equation}
    \hat{V}_v \psi = C_V c_{1/4} \sqrt{\Lambda_v} \int_{-\infty}^\infty e^{s/4} \Bar{A}_v (e^s \mathbb{I} + \Bar{A}_v)^{-1} \psi \, ds,
\end{equation}
with $c_{1/4} = 1/(\pi \sqrt{2})$. Now, choose a grid spacing $h > 0$, non-negative integers $K_-, K_+$ and nodes $s_k = kh, \tau_k = e^{kh}$ with $k = -K_-, \cdots, K_+$. Define the rational approximant
\begin{equation}
    R_{h, K_-, K_+}^{(\alpha)}(\Bar{A}) := c_\alpha h \sum_{k = -K_-}^{K_+} e^{\alpha kh}(\Bar{A} + e^{kh} \mathbb{I})^{-1} \Bar{A},
    \label{eq:R}
\end{equation}
now with $c_\alpha = \sin(\pi \alpha)/\pi$. For the vertex volume and $\alpha=1/4$, the shifted resolvent quadrature (SRQ) form of the matrix free volume operator takes the form
\begin{equation}
    \hat{V}_{v, h, K_-, K_+}^{\mathrm{SRQ}} := C_V \sqrt{\Lambda_v}R_{h, K_-, K_+}^{(1/4)}(\Bar{A}_v),
\end{equation}
or equivalently
\begin{equation}
    \hat{V}_{v, h, K_-, K_+}^{\mathrm{SRQ}} = C_V \sqrt{\Lambda_v} c_{1/4} h \sum_{k = -K_-}^{K_+} e^{kh/4} (Q_v^2 + \Lambda_v^2 e^{kh}\mathbb{I})^{-1} Q_v^2,
    \label{eq:srqvolumeexplicit}
\end{equation}
in unscaled variables. One can further show that such an expression indeed annihilates the volume kernel. For example, let $P_0$ be the orthogonal projector onto $\ker Q_v = \ker A_v = \ker \Bar{A}_v$. Each summand in \eqref{eq:R} has the form $(\Bar{A}_v + \tau \mathbb{I})^{-1}\Bar{A}_v$ with $\tau > 0$. Since $\Bar{A}_vP_0 = 0$, then such summands are also zero, and thus the finite sum is zero on $P_0 \Hcal_v$, in other words the exact zero volume vector is annihilated exactly at finite quadrature order (see \ref{sec:nearkernelstates} for more).

Additionally, the scalar function $r_\tau(\lambda) = \lambda/(\lambda + \tau)$ is real and non-negative for $\lambda \geq 0, \tau > 0$ and hence, $(\Bar{A}_v + \tau\mathbb{I})^{-1} \Bar{A}_v = r_\tau(\Bar{A}_v)$ is self-adjoint and positive semidefinite by the spectral theorem. All quadrature weights are positive, and thus the linear combination is positive and self-adjoint. The finite approximant is therefore itself a valid positive volume-like operator which can be implemented numerically as repeated action of oracles which do not materialise any dense matrix representations and allow for the application of volume operator in scalable numerical simulations.

\subsection{\label{subsec:shiftedsolves}Matrix free shifted resolvent system}

The presented SRQ expression \eqref{eq:srqvolumeexplicit} for the vertex volume provides a manner in which the vertex volume can be applied algorithmically by using $\ApplyQ_v$ instead of materialising $Q_v$. However, it also contains a set of shifted systems of the form $(\Bar{A} + \tau \mathbb{I})^{-1} \Bar{A}$ which need to be solved when applying the volume on a given state. This, however, is structurally similar to modern matrix free solvers used in neural quantum state variational Monte Carlo. Specifically, in stochastic reconfiguration (SR)~\cite{Sorella:1998,Sorella:2001,Carleo:2016svm}, one updates variational parameters $\theta$ by solving a damped linear system
\begin{equation}
    \label{eq:dampedsr}
    (S + \lambda \mathbb{I})^{-1} \delta \theta = -g
\end{equation}
where $S$ is the quantum geometric tensor (QGT)~\cite{Amari:1998,Sorella:2001} or covariance matrix of logarithmic derivatives of the
variational wavefunction. In neural quantum states, such an $S$ is often too large to form explicitly. If $O$ denotes the Jacobian-like map from parameter perturbations to logarithmic derivative samples, then schematically $S = O^\dagger O$. Modern implementations apply $S$ to a vector by first computing a Jacobian-vector product $Ox$ and then a vector-Jacobian product $O^\dagger (Ox)$, without forming $O$ or $S$~\cite{Baydin:2018,jax2018github}.

The present volume algorithm uses the same pattern. The positive operator whose fractional power is needed is $A_v$ and the shifted systems
\begin{equation}
    \label{eq:shiftedsystem}
    (\Bar{A}_v + \tau\mathbb{I})y = g,
\end{equation}
where $g = \Bar{A}_v \psi$, and $y$ is an approximate solution, are directly analogous to damped SR systems \eqref{eq:dampedsr}. The role of automatic differentiation VJPs in neural quantum states is played here by the adjoint recoupling action, which is already encoded by the Hermiticity of the grasping operator $G_{IJK}$. Thus, for $\tau > 0$, one solves the shifted equations \eqref{eq:shiftedsystem} without forming the matrix $\Bar{A}_v + \tau\mathbb{I}$ since a Krylov solver~\cite{HestenesStiefel:1952,Saad:2003} needs only the product
\begin{equation}\label{eq:shifted-matvec}
  p\mapsto(\bar A_v+\tau \mathbb{I})p
  =\frac{1}{\Lambda_v^2}\ApplyQ_v(\ApplyQ_v(p))+\tau p.
\end{equation}
Thus the only nontrivial black-box operation is $\ApplyQ_v$, whereby one can use the algorithms derived in, for example, \cite{Thiemann:1996at,Brunnemann:2004xi,brunnemann-and-rideout,BrunnemannRideout:2007xk} for such a case. Since $\bar A_v+\tau I\ge\tau I>0$, conjugate gradient methods are applicable~\cite{HestenesStiefel:1952,Saad:2003}. If the implementation uses complex arithmetic and the Hermitian convention for $Q_v$, inner products must be Hermitian. If the implementation uses the real antisymmetric convention, the system is real symmetric positive definite for $\Bar{A}_v=-q_v^2/\Lambda_v^2$.

Note that the SRQ expression contains \emph{multiple} shifted systems which are to be solved. An important observation is that all shifted systems share the same Krylov space $\mathcal{K}(\Bar{A}_v + \tau \mathbb{I}, g) = \mathcal{K}(\Bar{A}_v, g)$. Consequently, one can use multi-shift conjugate gradient methods which exploit this fact to update all shifted solutions using one Krylov process~\cite{Jegerlehner:1996,FrommerGlassner:1998,BaumannVanGijzen:2015}. The expensive operation per Krylov iteration is one application of $\Bar{A}_v$, hence two applications of $Q_v$.  The remaining operations are vector updates and scalar recurrences for the shifts. Shifted systems can also be solved independently and in parallel, which may be simpler on accelerators or distributed architectures.

%
%

\section{\label{sec:numerics}Numerical simulations}

The purpose of the numerical experiments presented here is not only to demonstrate scalability but rather, equally importantly, to validate the SRQ construction against an exact reference in regimes where such a reference can still be built. We therefore work on a fixed embedded $K_5$ graph.\footnote{The $K_5$ validation uses an embedding which produces, in the local leg order, the signs for $(123),(124),(134),(234)$ to be for $v_0:(+,-,+,+)$, $v_1:(-,+,-,+)$, $v_2:(+,-,-,-)$, $v_3:(-,+,+,+)$, and $v_4:(+,-,-,-)$.} The embedding enters only through the orientation signs $\epsilon_{IJK}(v)$ of triples of incident edge tangents. Throughout this section the overall dimensional prefactor is set to \(C_V=1\), so all reported numbers refer to the dimensionless Ashtekar-Lewandowski operator \(\sqrt{\abs{Q_v}}\), with \(Q_v\) defined above.

In the implementation study we tested several equivalent realizations of the same finite SRQ operator. Specifically, (i) a spectral evaluation of the SRQ scalar function on the dense local \(Q_v\) blocks, (ii) shifted linear solves using dense local applications of \(Q_v\), and (iii) shifted linear solves using a recoupling-action oracle for \(Q_v\). On the 4-valent vertices of \(K_5\), gauge invariance reduces the Ashtekar-Lewandowski density to the standard Brunnemann-Thiemann 4-valent action of the \(q_{123}\) grasping \cite{Brunnemann:2004xi}, multiplied by the reduced orientation sign. The figures reported below use this Brunnemann-Thiemann 4-valent action as the \(Q_v\) oracle. All mentioned alternative SRQ realizations were used as internal consistency checks and agree with it at roundoff in the tested sectors.

The cutoff used below is a doubled-spin cutoff.  For \(j_{\max}\in\{1,2\}\), each edge spin is restricted by $0 \leq 2j_e \leq j_{\max}$. Thus $2j_{\max}=1$ contains spins $j=0,\tfrac12$, while $j_{\max}=2$ contains $j=0,\tfrac{1}{2},1$. The spin network basis is the gauge invariant recoupling basis
\begin{equation}
    \ket{\sigma}
    =
    \ket{\{j_e\}_{e\in E(K_5)},\{a_v\}_{v\in V(K_5)}} ,
\end{equation}
where $a_v$ is the single intermediate recoupling spin at the 4-valent vertex $v$. A basis label is included only when the Clebsch-Gordan admissibility conditions are satisfied at every vertex. This gives $\dim\Hcal_{j_{\max}=1}(K_5)=140$ and $\dim\Hcal_{j_{\max}=2}(K_5)=10989$.

For the exact reference, $Q_v$ is constructed independently in each fixed incident-spin sector $\bm j_v=(j_1,j_2,j_3,j_4)$. We first build the orthonormal 4-valent intertwiner basis from Clebsch-Gordan coefficients in the magnetic basis. In that basis the triple grasping is evaluated through the
commutator form~\cite{DePietri:1996tvo,Brunnemann:2004xi}
\begin{equation}
    q^{\rm comm}_{IJK}
    =
    \left[
      (\bm X_I+\bm X_J)^2,
      (\bm X_J+\bm X_K)^2
    \right],
\end{equation}
projected to the gauge invariant intertwiner space. With the orientation signs of the fixed \(K_5\) embedding, the Hermitian local density is
\begin{equation}
    Q_v(\bm j_v)
    =
    \ii\sum_{I<J<K}\epsilon_{IJK}(v)\,
    q^{\rm comm}_{IJK}(\bm j_v).
\end{equation}
The exact local Ashtekar-Lewandowski volume block is then obtained by a literal spectral square root, $V_v^{\mathrm{AL}}(\bm j_v) = \sqrt{\abs{Q_v(\bm j_v)}}$.

Finally these exact blocks are embedded into the full $K_5$ basis by acting only on the recoupling label $a_v$ and leaving all edge spins and all other vertex intertwiners fixed.  For cutoff $j_{\max}=1$, we also materialised the full dense $140\times140$ graph operator and checked the local-block embedding against it. The relative action error on the trial state was $1.383\times10^{-16}$. For cutoff $j_{\max}=2$, a full dense graph matrix is not materialised. Rather, the reference is still exact but exactness is kept locally by dense diagonalisation of the finite vertex-sector blocks.

In the small-cutoff validation runs, all expectation values are computed by full summation over the finite spin network basis. For an operator $O$,
\begin{equation}\label{eq:numerical-expectation}
    \langle O\rangle_\psi
    =
    \frac{\langle\psi,O\psi\rangle}{\langle\psi,\psi\rangle}
    =
    \sum_{\sigma}
    \frac{\abs{\psi_\sigma}^2}{\norm{\psi}^2}
    \frac{(O\psi)_\sigma}{\psi_\sigma}.
\end{equation}
Therefore, no Monte Carlo estimator is used in these small-cutoff validation figures. The trial state $\psi$ is a fixed non-vanishing complex vector on the exact basis, with mild spin-dependent amplitudes and nontrivial phases. Its role is only to probe the operator action without introducing symmetry-protected cancellations. The exact total-volume expectations are
\begin{equation}
    \left\langle\sum_v V_v\right\rangle_\psi
    =
    6.104758328461
    \quad(j_{\max}=1),
    \qquad
    7.968227239035
    \quad(j_{\max}=2),
\end{equation}
which act as the reference exact numerical values for the following small-cutoff consistency and validity checks.

\subsection{\label{subsec:numerics-M}Convergence with the number of shifted systems}

\begin{figure}[ht]
\centering
\begin{minipage}{0.48\textwidth}
  \centering
  \includegraphics[width=\linewidth]{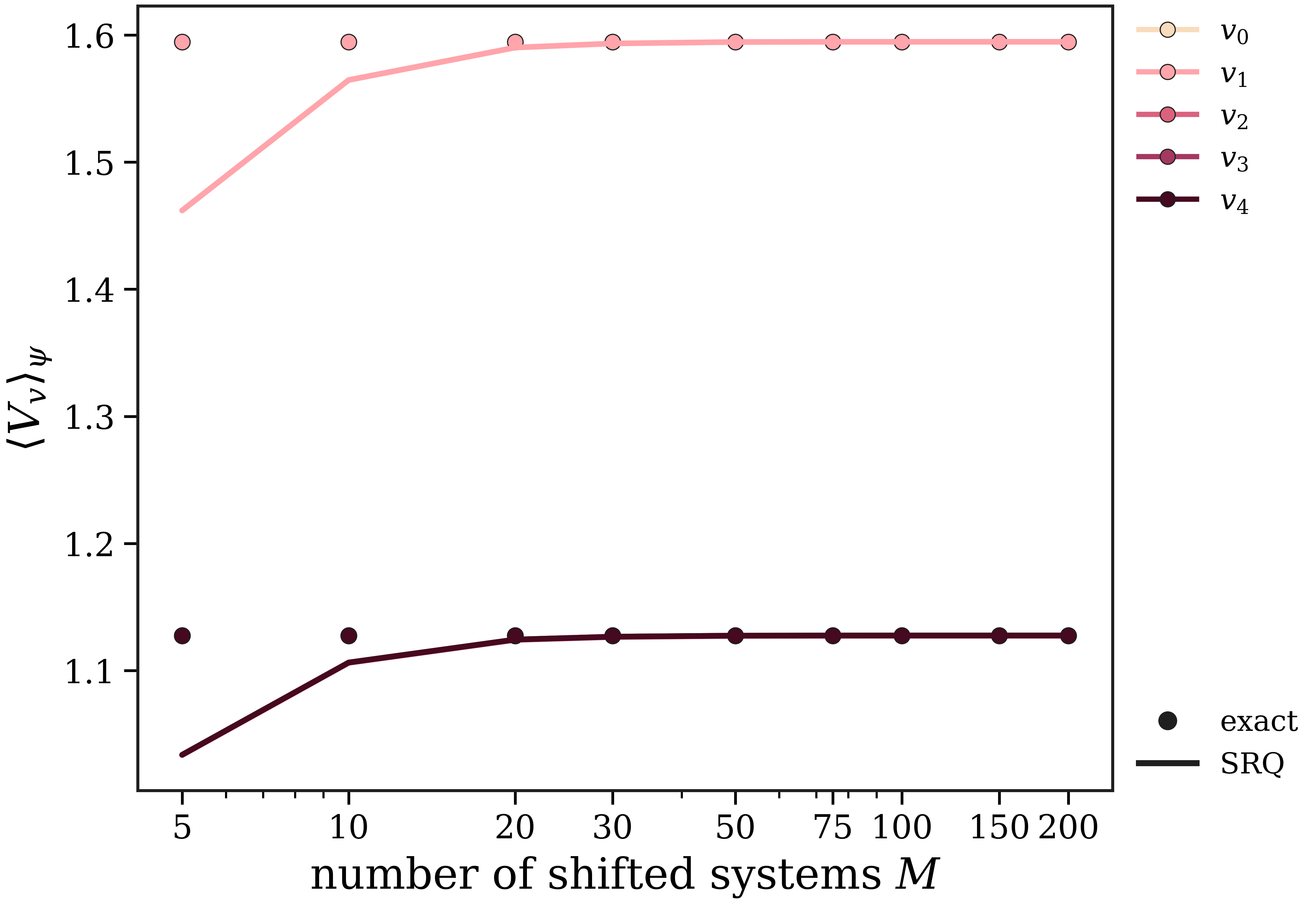}
\end{minipage}
\hfill
\begin{minipage}{0.48\textwidth}
  \centering
  \includegraphics[width=\linewidth]{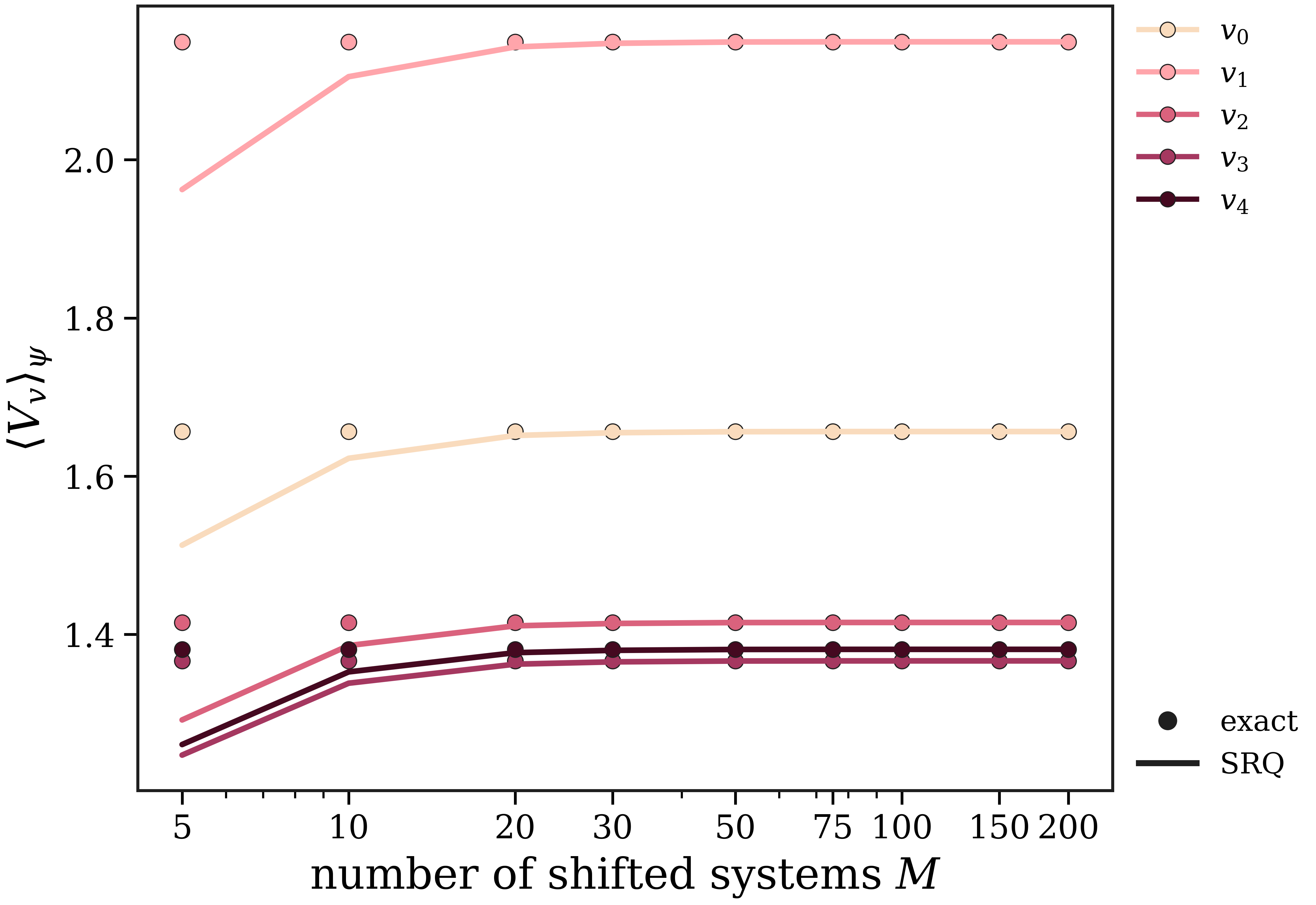}
\end{minipage}
\caption{\label{fig:k5-vertex-M-convergence}
Vertex-resolved convergence of the SRQ approximation on the fixed $K_5$ graph. The left panel shows doubled-spin cutoff $j_{\max}=1$ and the right panel shows $j_{\max}=2$. Exact dense-block values of $\langle V_v\rangle_\psi$ are shown as dots, while the SRQ values with sector-wise analytic $\Lambda_{\rm an}$ are shown as continuous curves. The rapid approach of the curves to the dots is the numerical manifestation of the root-exponential convergence expected from the logarithmic shifted-resolvent quadrature.}
\end{figure}

We first fix the sector-wise analytic bound $\Lambda_v=\Lambda_{\rm an}$ (see \ref{sec:choosinglambda}) and vary the number $M = K_- + K_+ + 1$ of shifted systems in the SRQ approximation. For each vertex, we compare $\langle V_v\rangle_\psi$ with $\langle V^{\mathrm{SRQ}}_{v,M}\rangle_\psi$. \Cref{fig:k5-vertex-M-convergence} shows the exact values as dots and the SRQ values as continuous curves.  The convergence is rapid and uniform across the five vertices. Note that for cutoff $j_{\max}=1$, several vertices have the same vertex volume expectation in this chosen state, and hence appear overlapped in the displayed figure. For cutoff $j_{\max}=1$, the largest absolute vertex error is $1.327\times10^{-1}$ at $M=5$, which quickly goes down as $M$ is increased to 200 to a value of $9.665\times10^{-9}$. Similarly, for cutoff $j_{\max}=2$, the corresponding errors are $1.867\times10^{-1}$ and $1.825\times10^{-8}$ at $M=5$ and $M=200$ respectively. Note that these numbers are vertex-wise errors, not errors in the total volume, and therefore give a direct test of the local operator action.

\begin{figure}[ht]
\centering
  \includegraphics[width=0.62\linewidth]{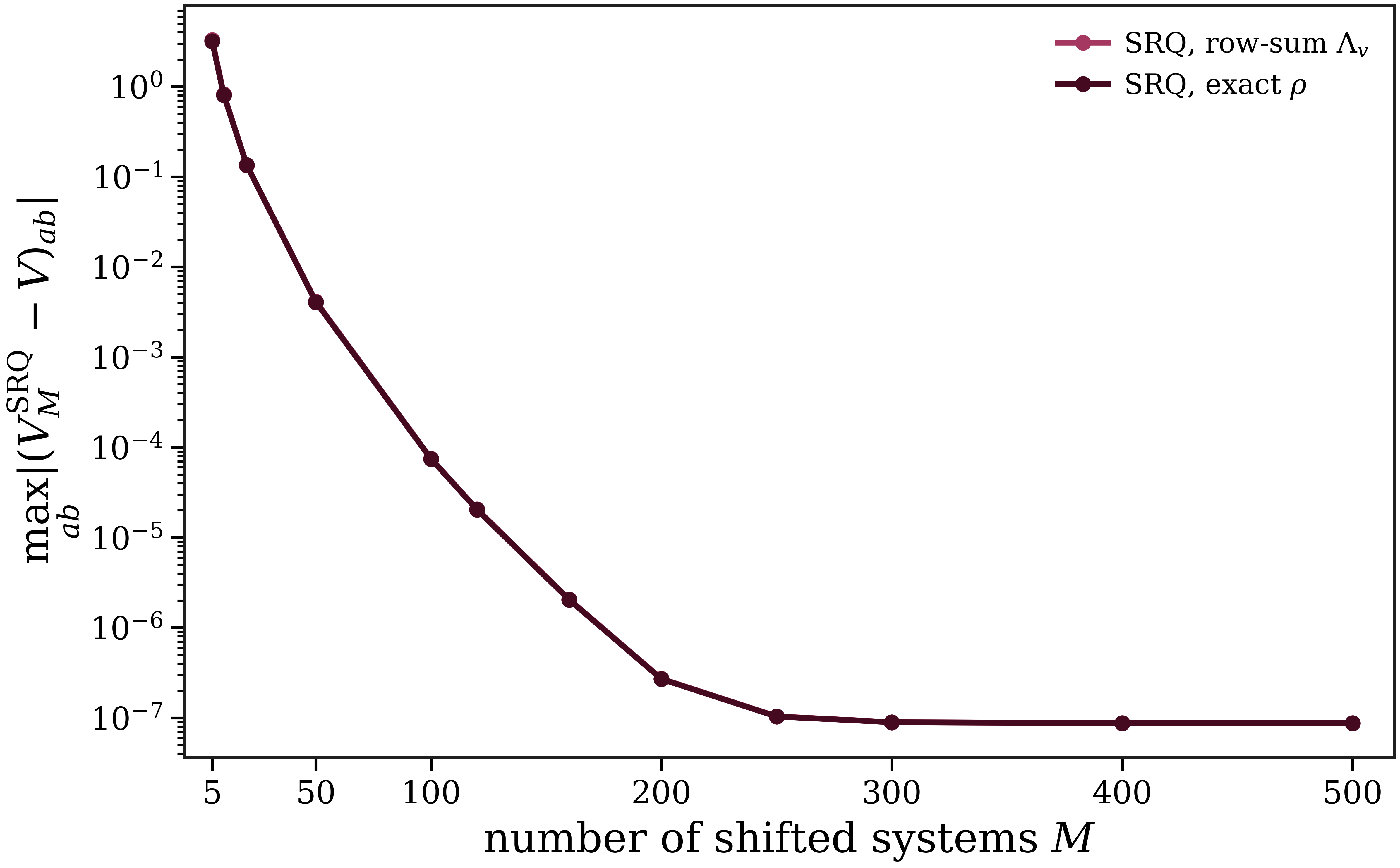}
\caption{\label{fig:four-valent-dense-error}
Entrywise dense-block validation of the SRQ volume action in a homogeneous 4-valent sector with $2j_i=20$ and $d_v=21$. The horizontal axis is the number of shifted systems $M$ on a linear scale. The vertical axis is the maximum absolute matrix-element error against the exact dense spectral square root $V_v=\sqrt{\abs{Q_v}}$, displayed on a logarithmic scale. The SRQ block is assembled by applying the matrix free action to canonical basis vectors. The row-sum bound $\Lambda_v=3.552858037\times10^3$ and the diagnostic exact-radius choice $\rho_v=3.320858159\times10^3$ give nearly identical convergence, reaching the $10^{-7}$ numerical floor by $M\simeq300$.}
\end{figure}

The preceding test is formulated at the level of expectation values in a fixed trial state. This is the relevant quantity for variational and Monte Carlo applications, but it is still a state-dependent contraction of the local operator. To test the finite SRQ action more directly, we also compare the matrix elements of the local volume block itself in a homogeneous 4-valent sector with $2j_1=\cdots=2j_4=20$. This sector has $d_v=21$ admissible intertwiners. We form the dense reference $V_v=\sqrt{\abs{Q_v}}$ by diagonalising the dense $Q_v$ block, and assemble $V^{\rm SRQ}_{v,M}$ by applying the SRQ action to each canonical intertwiner basis vector. We then compute
\begin{equation}
    E_M^{\rm max}
    =
    \max_{a,b}
    \abs{
      \left(V^{\rm SRQ}_{v,M}-V_v\right)_{ab}
    }.
\end{equation}
In \Cref{fig:four-valent-dense-error}, the horizontal axis is the number of shifted systems $M$ on a linear scale, while the vertical axis is $E_M^{\rm max}$ on a logarithmic scale. Further, we choose two different ways of choosing the scaling parameter $\Lambda_v$ (see \ref{sec:choosinglambda}), one derived explicitly from $\rho_v = \norm{Q_v}$, and one estimated without the need to materialise $Q_v$ to demonstrate the effect of loosely choosing such a parameter.

For the row-sum bound, the maximum absolute matrix-element error decreases from $3.276$ at $M=5$ to $4.133\times10^{-3}$ at $M=50$, and rapidly towards $1.038\times10^{-7}$ at $M=250$. Increasing to $M=300,400,500$ does not materially improve the result, with the error saturating near $8.7\times10^{-8}$. The near coincidence with the exact-$\rho_v$ diagnostic curve indicates that the scalable row-sum bound is already sufficiently tight for this sector. The remaining plateau is dominated by numerical precision in the dense reference and SRQ assembly rather than by the choice of $\Lambda_v$.
The scale of the decay agrees with the explicit estimate in \ref{sec:errorbounds}. For the row-sum scale in this sector, the bound $\sqrt{\Lambda_v}\epsilon_M$, with $\epsilon_M=E_{\rm disc}+E_{\rm left}+E_{\rm right}$ from \eqref{eq:operator-error-bound-general}, gives $1.370\times10^{-2}$, $2.563\times10^{-4}$, and $9.140\times10^{-7}$ at $M=50,100,200$, while the corresponding measured entry-wise errors are $4.133\times10^{-3}$, $7.478\times10^{-5}$, and $2.711\times10^{-7}$. Thus the pre-plateau data lie safely below the predicted operator-norm bound and follow the same root-exponential scale, while the later saturation begins once the theoretical quadrature error has fallen below the numerical floor of the assembled validation block.

The important point is that the convergence is observed at the level of the individual vertex operators, not only after summing over vertices, and also at the level of the local matrix elements in an independent homogeneous sector. We note that this is a stronger validation of the construction, since the total volume could in principle hide local errors by cancellation. In the present data, every vertex expectation approaches its dense-block reference with the same systematic behaviour, and the cutoff $j_{\max}=2$ case shows no qualitative degradation despite the larger spin content and the relatively larger graph Hilbert space.

\subsection{\label{subsec:numerics-highcutoff}High-cutoff matrix free Monte Carlo}

Aside from the small-spin validation tests above, the same algorithm can be used to compute volume expectation values in graph sectors far outside the range where exact dense methods are available. To demonstrate this, we again use the fixed embedded $K_5$ graph, but now work at doubled-spin cutoff $j_{\max}=250000$, equivalently $j_e\leq125000$. We fix the ten edge labels, in the graph edge order used by the Hilbert-space construction, to
\begin{equation}
    \begin{split}
    (2j_e)_{e=1}^{10}
    =
    (250000,248000,246000,&244000,242000,\\
     &240000,238000,236000,234000,232000).
    \end{split}
\end{equation}
The remaining degrees of freedom are the five 4-valent recoupling channels. Thus a basis state in this fixed-edge sector has the form $\sigma=(\bm j,a_0,a_1,a_2,a_3,a_4)$, where $\bm j$ denotes the fixed edge labels and $a_v$ is an admissible intermediate doubled spin at vertex $v$. Since the edge labels are fixed, admissibility factorizes over the five vertex channels. The local intertwiner dimensions are $\big(\abs{\mathcal A_0},\ldots,\abs{\mathcal A_4}\big) = (124001,121001,122001,122001,122001)$, and hence the fixed-edge graph sector has $\abs{\mathcal S_{\bm j}} = \prod_{v=0}^{4}\abs{\mathcal A_v(\bm j)} \approx 2.7246 \times 10^{25}$ basis states. Materialising dense graph operator on this sector is plainly meaningless. More importantly for the local volume problem, the largest single vertex block has dimension $124001$. This example is therefore outside the dense local-block strategy itself, not merely outside full graph-level enumeration\footnote{Materialising only one dense complex matrix for $Q_v$ on that block would require $124001^2$ complex entries, which is at least $246$ GB before forming $Q_v^2$, $V_v$, eigenvectors, or any diagonalisation workspace.}.

\begin{figure}[ht]
\centering
\begin{minipage}{0.48\textwidth}
  \centering
  \includegraphics[width=\linewidth]{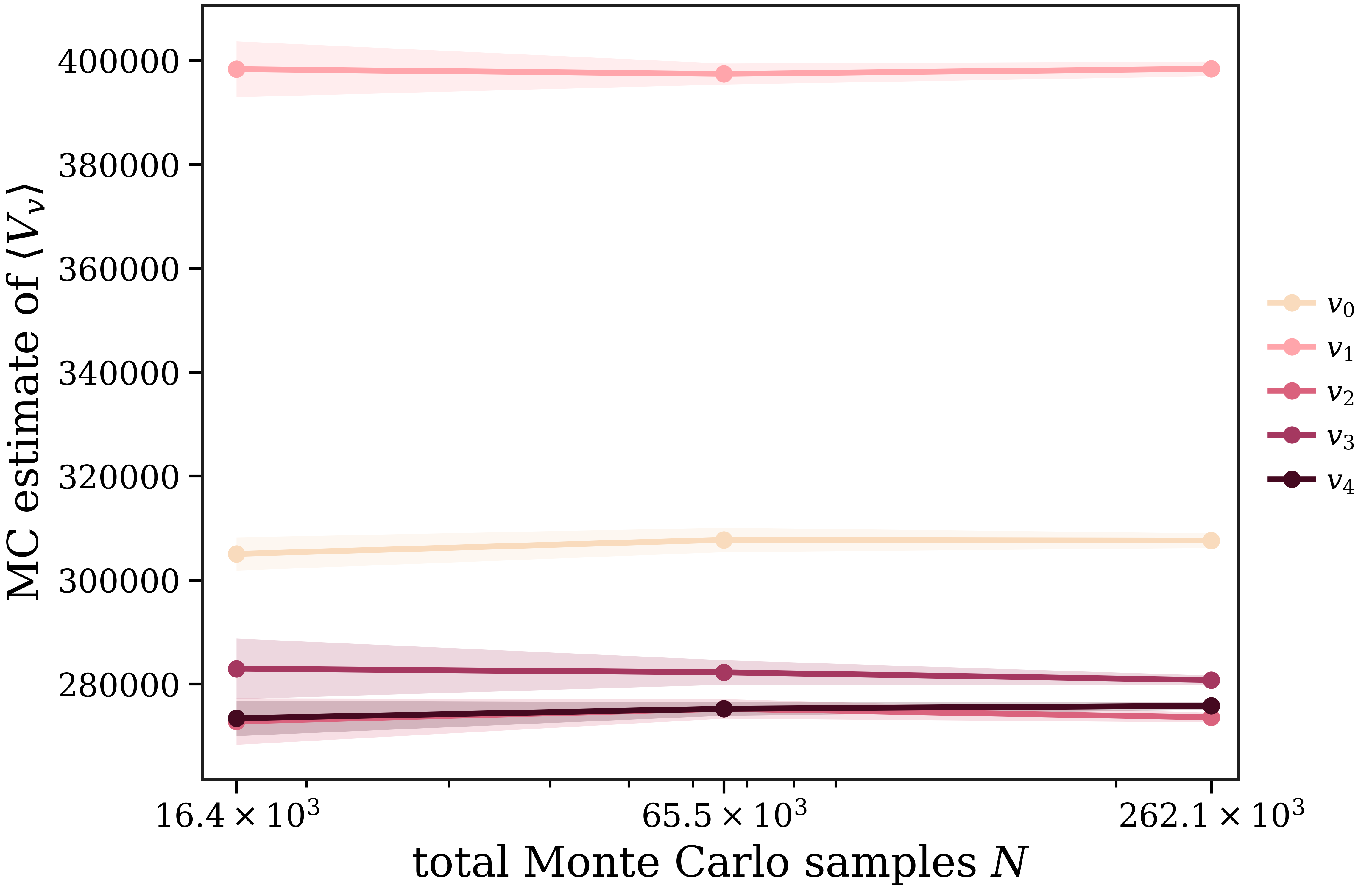}
\end{minipage}
\hfill
\begin{minipage}{0.48\textwidth}
  \centering
  \includegraphics[width=\linewidth]{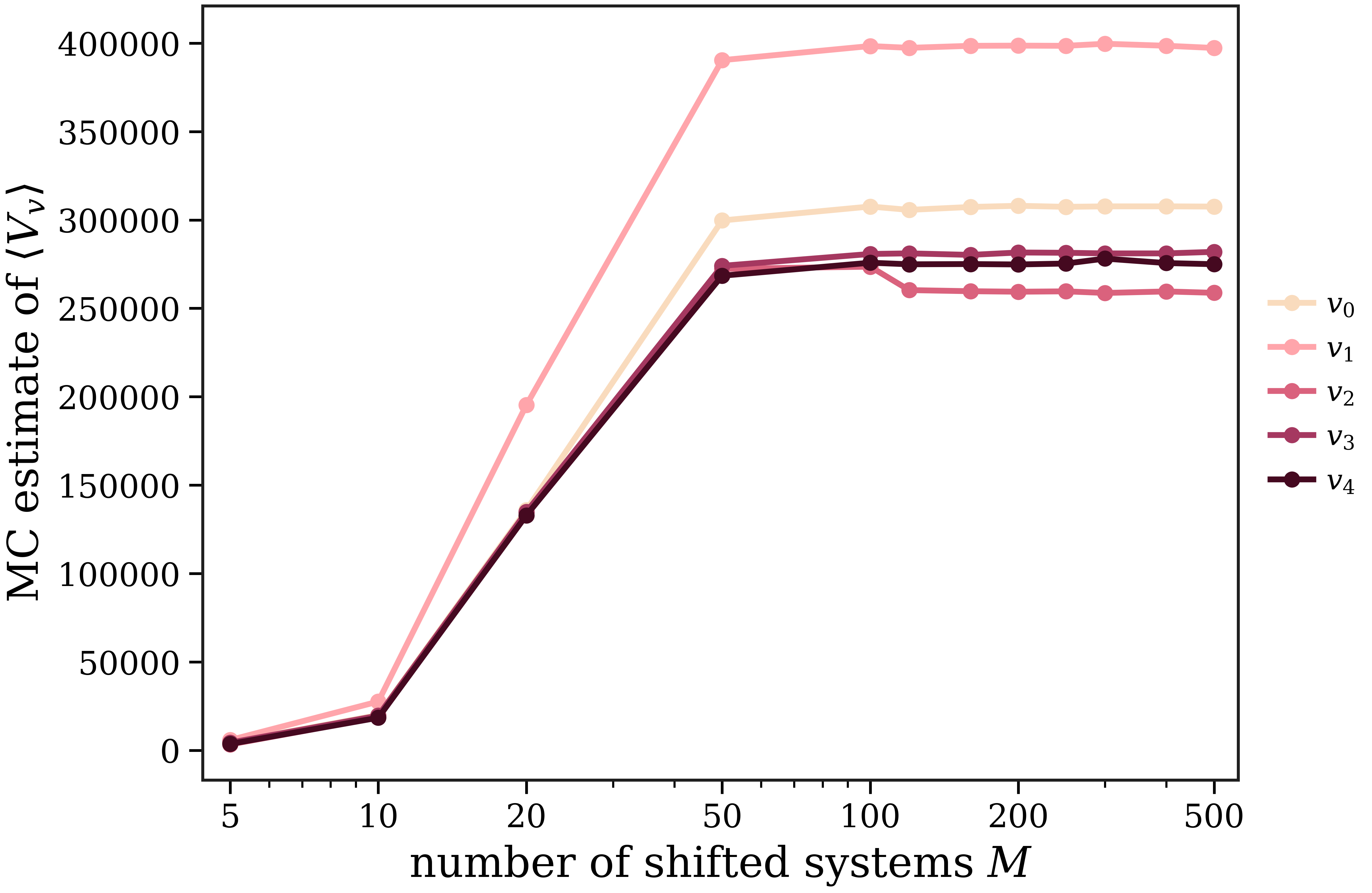}
\end{minipage}
\caption{\label{fig:k5-highcutoff-mc}
Matrix free SRQ Monte Carlo estimates of high-cutoff $K_5$ vertex-volume expectations. The fixed-edge sector has doubled-spin cutoff $j_{\max}=250000$ and dimension $2.7246 \times 10^{25}$. The local vertex dimensions are $(124001,121001,122001,122001,122001)$, thus dense local $Q_v$ matrix materialisation is infeasible. Left panel: sample-count dependence at fixed number of shifted systems $M=100$. Right panel: dependence on the number of shifted systems at fixed total sample count $N=262144$. Shaded regions are one standard error from 16 independent sample chains.}
\end{figure}

For this high-cutoff test we take the trial state to be the uniform-amplitude state on the fixed-edge sector, $\psi_{\rm unif}(\sigma)=1$ with $\sigma\in\mathcal S_{\bm j}$. The expectation value can then be written in local-estimator form as
\begin{equation}
    \langle V_v\rangle_{\psi_{\rm unif}}
    =
    \frac{1}{\abs{\mathcal S_{\bm j}}}
    \sum_{\sigma\in\mathcal S_{\bm j}}
    \frac{(V_v\psi_{\rm unif})(\sigma)}{\psi_{\rm unif}(\sigma)}.
\end{equation}
In the sample-count scan below, $V_v$ is replaced by the finite SRQ action $V^{\rm SRQ}_{v,M}$ with $M=100$. To separate Monte Carlo sampling error from finite-quadrature error, we also keep the total sample count fixed at $N=262144$ and evaluate the same estimator at $M=5,10,20,50,100,120,160,200,250,300,400,500$. For a sampled state $\sigma=(\bm j,a_0,\ldots,a_4)$, the local vector on which the vertex operator acts is the all-ones vector on the admissible channel set $\mathcal A_v(\bm j)$. The sampled local estimator is therefore
\begin{equation}
    \ell_{v,M}(\sigma)
    =
    \left[
      V^{\rm SRQ}_{v,M}(\bm j_v)\,\mathbf 1_v
    \right]_{a_v},
\end{equation}
and the Monte Carlo estimate is
\begin{equation}
    \widehat{\langle V_v\rangle}_{M,N}
    =
    \frac{1}{N}\sum_{n=1}^{N}\ell_{v,M}(\sigma_n).
\end{equation}
Here the samples $\sigma_n$ are independent uniform samples from the factorized product of the five admissible recoupling-channel sets in the fixed-edge sector. The error bands in \Cref{fig:k5-highcutoff-mc} are one standard error estimated from 16 independent batches. 

\Cref{fig:k5-highcutoff-mc} shows that the matrix free estimator remains directly computable in this regime. The left panel shows that increasing the number of samples narrows the Monte Carlo uncertainty without changing the finite-$M$ SRQ operator being sampled. The right panel shows the complementary finite-$M$ behaviour over the same range of shifted-system counts used in \Cref{fig:four-valent-dense-error}. The very low orders $M=5,10,20$ are not yet in the asymptotic regime, while the estimates from $M=50$ onward are much closer to stable values. The remaining movement between $M=100$ and $M=120$, most visibly for $v_2$, is larger than the displayed Monte Carlo standard error and should therefore be interpreted as finite-SRQ bias rather than sampling noise. For $M\geq120$, the estimates remain essentially stable through $M=500$ at the resolution of the displayed Monte Carlo uncertainty. In the reported local run, despite a performance suboptimal implementation, all five vertex estimators at the three displayed sample counts, together with metadata construction and output generation, required on average $1.52$ seconds of measured end-to-end time on a standard Apple M5 silicon chip. The fixed-sample sweep through $M=500$ required $18.8$ seconds in the same local setup. While these timings are implementation- and hardware-dependent, they nevertheless show that an inaccessible regime to dense materialisation is not only accessible using the SRQ but also reasonably efficient, as the important point is instead the scaling structure as the computation depends on local banded recoupling data, the number of shifted systems, and the number of samples, not on materialising a $2.7\times10^{25}$ dimensional graph space or a $124001\times124001$ dense local volume density matrix.

\subsection{\label{subsec:numerics-lambda}Sensitivity to the choice of \(\Lambda_v\)}

We next keep $M=20$ fixed and vary the sector-wise upper bound $\Lambda_v$. The value $M=20$ is chosen as intentionally modest, given that at $M=50$ and above the curves are already so close to the exact values that the dependence on $\Lambda_v$ is barely visible on the scale of the expectation values. The analytic bound used in the previous scan, and denoted by $\Lambda_{\rm an}$, is determined without forming or diagonalising $Q_v$. For each fixed 4-valent incident-spin sector $\bm j_v=(j_1,j_2,j_3,j_4)$, we use the product-norm estimate of \ref{sec:choosinglambda}. In the 4-valent gauge-invariant case this reduces to
\begin{equation}\label{eq:numerics-analytic-lambda}
    \Lambda_{\rm an}(v,\bm j_v)
    =
    6\abs{\sigma_v}
    \sqrt{
      j_1(j_1+1)
      j_2(j_2+1)
      j_3(j_3+1)
    },
\end{equation}
where $\sigma_v = \epsilon_{012}(v) - \epsilon_{013}(v) + \epsilon_{023}(v) - \epsilon_{123}(v)$
is the reduced orientation sign for the chosen local ordering of the four incident edges. This is precisely the sector-wise a priori enclosure used in the SRQ action below. We compare diagnostic choices proportional to the exact sector radius $\rho_v=\norm{Q_v}$, namely $\rho_v, 1.05\rho_v, 1.25\rho_v, 1.5\rho_v, 2\rho_v$, with the analytic bound $\Lambda_{\rm an}$ and the looser bounds $1.25\Lambda_{\rm an}$ and $2\Lambda_{\rm an}$. We note that the radius-based choices are included here as diagnostics for how much accuracy is lost when the admissible bound is made looser and not to be confused as a scalable method for computing the $\Lambda_v$ bound (see \ref{sec:choosinglambda}).

\begin{figure}[ht]
\centering
\begin{minipage}{0.48\textwidth}
  \centering
  \includegraphics[width=\linewidth]{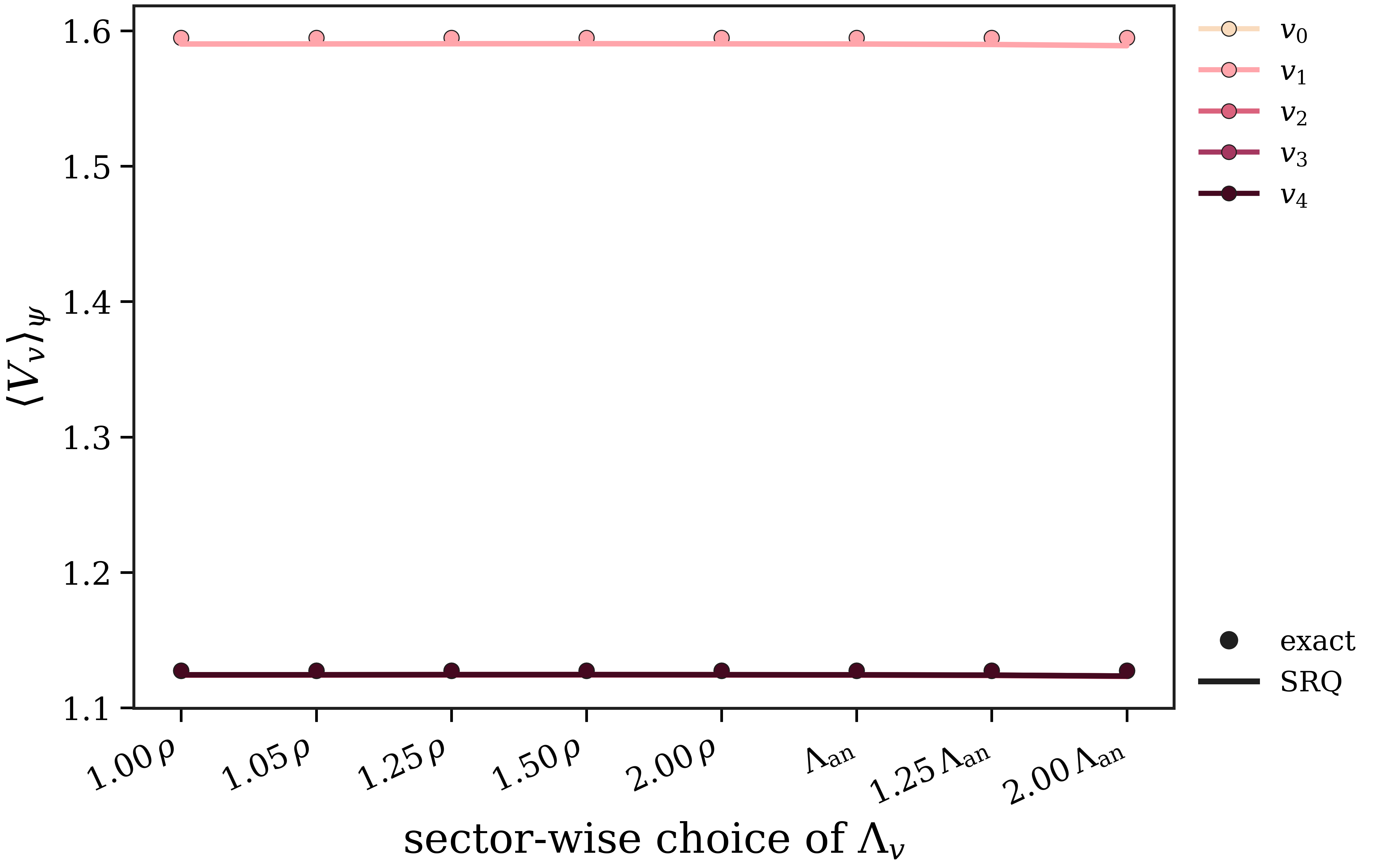}
\end{minipage}
\hfill
\begin{minipage}{0.48\textwidth}
  \centering
  \includegraphics[width=\linewidth]{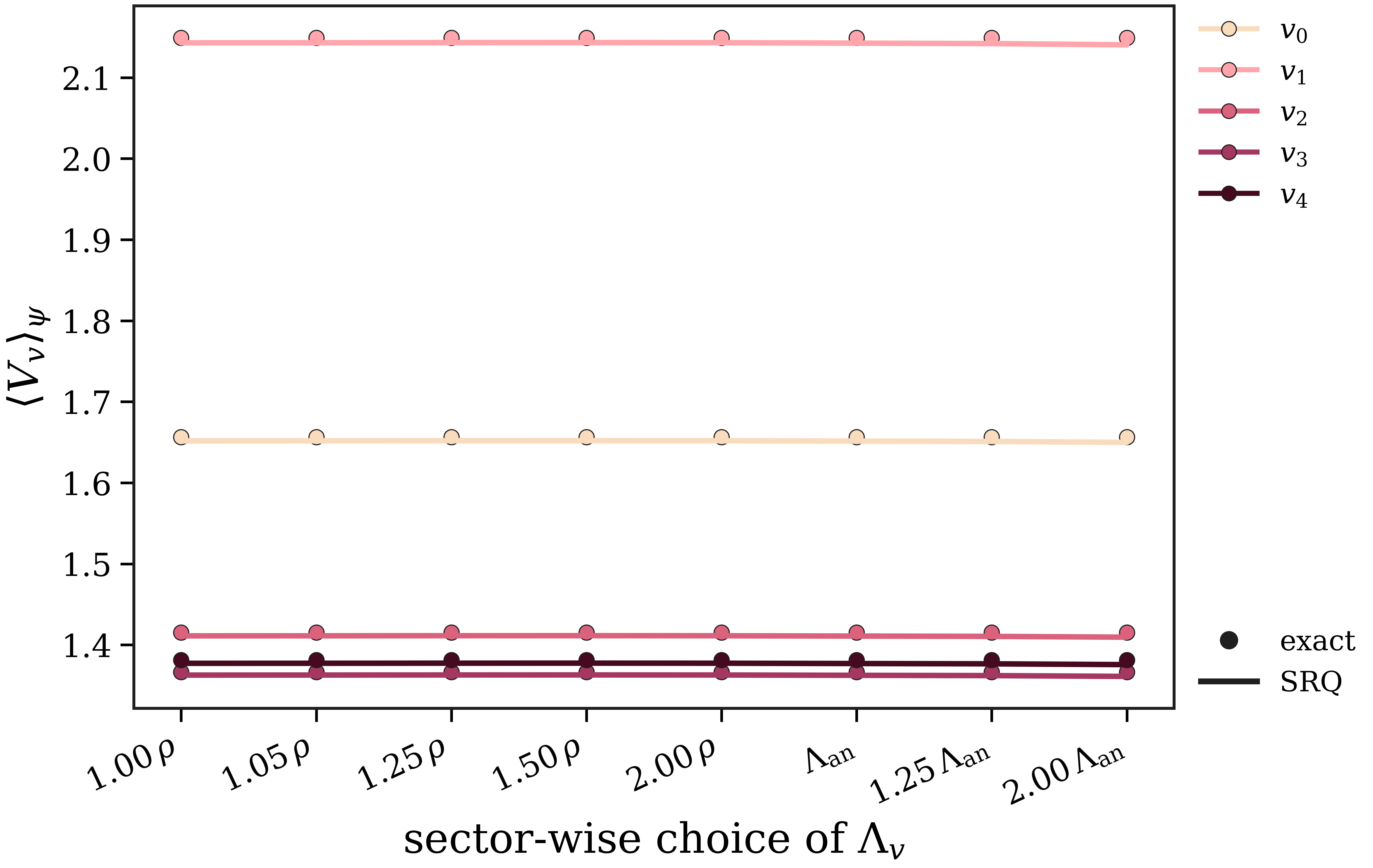}
\end{minipage}
\caption{\label{fig:k5-vertex-lambda-sensitivity}
Sensitivity of the Ashtekar-Lewandowski vertex-volume expectation values to the sector-wise bound $\Lambda_v$, at fixed SRQ order $M=20$ for $j_{\max}=1$. The exact dense-block expectations are shown as dots and the SRQ estimates as continuous curves. The analytic bound $\Lambda_{\rm an}$ is already close to these diagnostic choices, while deliberately looser choices increase the finite-$M$ bias without changing the limiting operator.}
\end{figure}

\Cref{fig:k5-vertex-lambda-sensitivity} shows that the dependence on $\Lambda_v$ is mild but systematic. For $j_{\max}=1$, the largest vertex error over the scan ranges from $4.294\times10^{-3}$ for $1.5\rho_v$ to $5.805\times10^{-3}$ for $2\Lambda_{\rm an}$, with the analytic bound itself giving $4.523\times10^{-3}$. For $j_{\max}=2$, the corresponding range is $5.786\times10^{-3}$ to $8.565\times10^{-3}$, with $\Lambda_{\rm an}$ giving $6.505\times10^{-3}$. This behaviour is consistent with \eqref{eq:relative-error-Lambda}. Namely, the algorithm remains faithful for any valid upper bound, while tighter sector-wise bounds reduce the prefactor in the approximation error.

This scan also clarifies the role of $\Lambda_v$ in practice. It is not a variational parameter and it is not tuned to fit the exact volume but rather it is a sector-wise spectral enclosure required to place $Q_v^2/\Lambda_v^2$ in the unit interval. Once this condition is satisfied, changing $\Lambda_v$
changes only the finite-order quadrature error. The analytic bound used in the scalable algorithm is therefore adequate for the present tests, while a tighter matrix free enclosure can be used when one wants to reduce the quadrature order needed for a target accuracy.

\subsection{\label{subsec:numerics-kernel}Kernel preservation}

\begin{figure}[ht]
\centering
\begin{minipage}{0.48\textwidth}
  \centering
  \includegraphics[width=\linewidth]{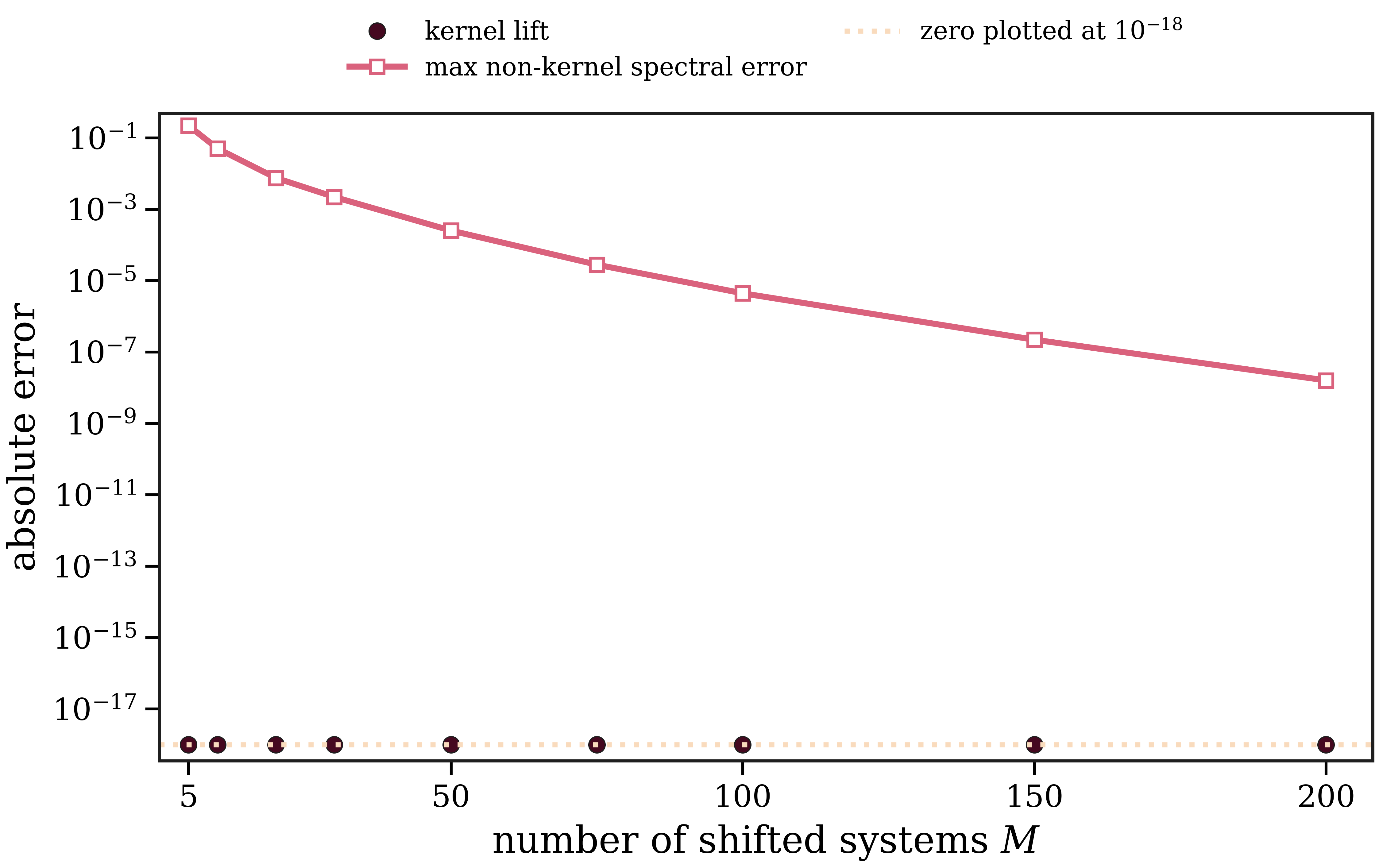}
\end{minipage}
\hfill
\begin{minipage}{0.48\textwidth}
  \centering
  \includegraphics[width=\linewidth]{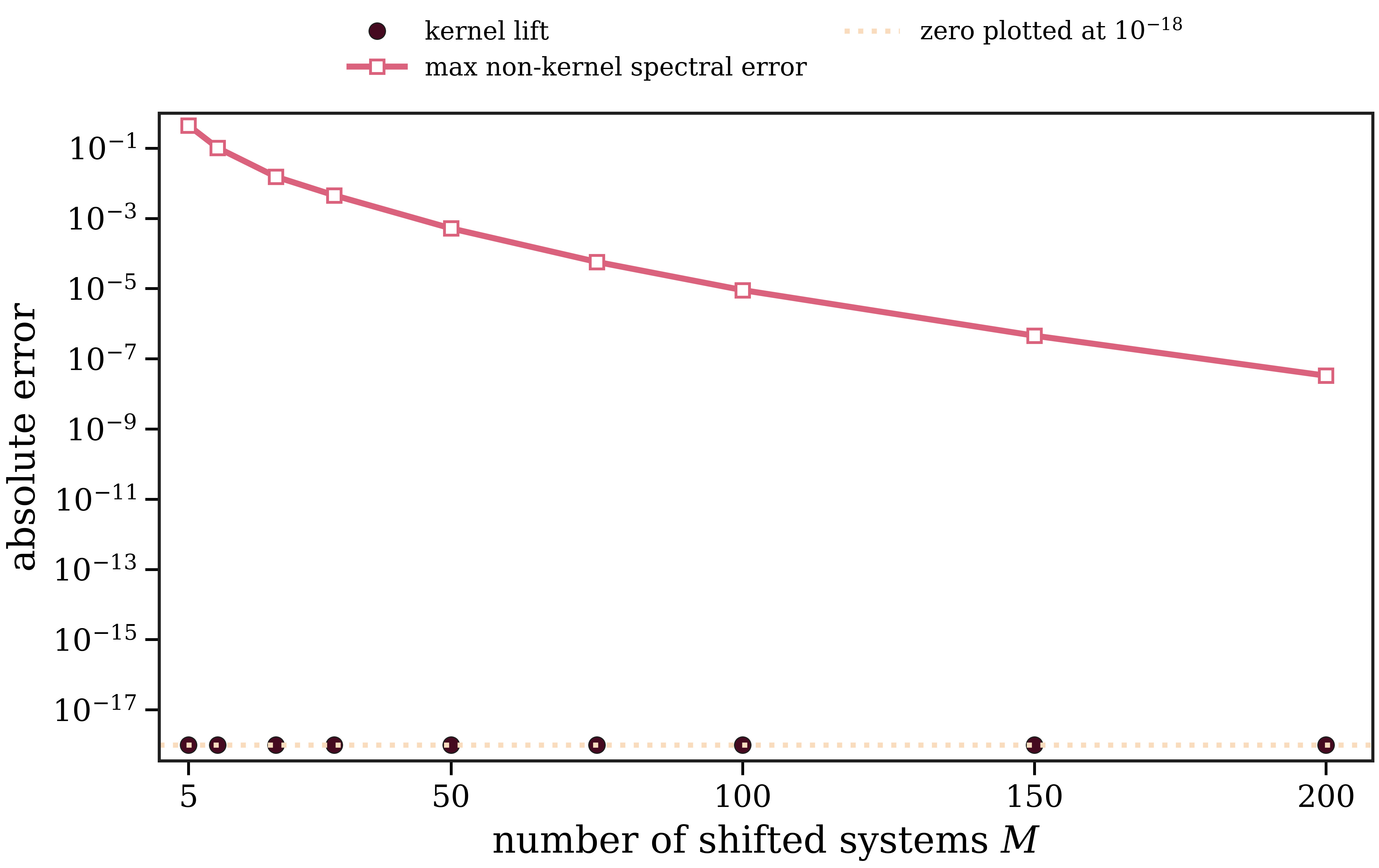}
\end{minipage}
\caption{\label{fig:k5-kernel-preservation}
Kernel preservation of the SRQ vertex-volume approximation on $K_5$. The kernel lift $L_M$ is zero for every tested quadrature order and is displayed at a fixed plotting floor of $10^{-18}$. The non-kernel spectral error is shown on the same axis to demonstrate that the test is nontrivial, in the sense that away from the kernel the SRQ approximation converges rapidly with $M$, while the kernel
is preserved exactly at finite order.}
\end{figure}

A central structural property of the SRQ formula is that it preserves the kernel of $Q_v$ exactly at \emph{finite} quadrature order. In the scaled form, every summand contains $(\bar A_v+\tau \mathbb{I})^{-1}\bar A_v$, with $\bar A_v=Q_v^2/\Lambda_v^2$. Hence if $u\in\ker Q_v$, then $\bar A_v u=0$ and every summand vanishes. We test this property directly in the same exact local sectors used above. For each sector, let $P_0^{(v,\bm j)}$ denote the projector onto $\ker Q_v(\bm j)$. We compute the kernel lift
\begin{equation}
    L_M
    :=
    \max_{v,\bm j}
    \norm{
      V^{\mathrm{SRQ}}_{v,M}(\bm j) P_0^{(v,\bm j)}
    }_2,
\end{equation}
and compare it with the maximum spectral error on the complement of the kernel. The result is shown in \Cref{fig:k5-kernel-preservation}. 

As shown in \Cref{fig:k5-kernel-preservation}, the kernel lift is exactly zero for all tested $M$ in both cutoffs, and the zero values are drawn at $10^{-18}$ only to make them visible on the logarithmic axis. In contrast, the non-kernel spectral error decreases from $2.191\times10^{-1}$
to $1.595\times10^{-8}$ for $j_{\max}=1$, and from $4.456\times10^{-1}$ to $3.289\times10^{-8}$ for $j_{\max}=2$, over the same range of $M$. The embedded local-sector count of kernel modes is 430 for $j_{\max}=1$ and 26555 for $j_{\max}=2$. Thus the algorithm is not converging to the correct kernel only in the limit, it annihilates the exact kernel throughout the finite-order approximation.

The series of tests, although some conducted at small spin cutoffs to validate against exact computations, isolate the three numerical features needed for a matrix free volume action. First, at fixed analytic sector bound the SRQ operator converges rapidly to the exact dense-block vertex volume. Second, the dependence on $\Lambda_v$ is controlled, looser valid bounds increase the finite-$M$ bias but do not change the limiting operator. Third, and most importantly for applications to constraint solving, the exact zero-volume subspace is not lifted by the finite quadrature approximation. The experiments therefore support the central use of SRQ as an algorithmic definition of the vertex volume action in settings where dense diagonalisation is unavailable, as shown in the high spin Monte Carlo example.

\subsection{\label{subsec:spectralanalysis}Spectral analysis}

The previous tests validate the action of the vertex volume on prescribed states and demonstrate the scalability of such a formulation. A complementary question is whether the same action can be used to probe spectral information of the volume operator without first constructing a dense volume matrix. This is natural in LQG, where the spectral distribution of higher-valent volume blocks has long been studied by exact diagonalisation in finite recoupling sectors \cite{brunnemann-and-rideout,BrunnemannRideout:2007xk}. The point of the present study is more modest and more algorithmic, namely we ask whether the finite SRQ volume operator itself can be supplied as a matrix free operator to a spectral estimator.

The spectral estimator used here is stochastic Lanczos quadrature (SLQ). Let $B$ be a finite-dimensional self-adjoint operator on a space of dimension $d$. Its normalised spectral measure is $\mu_B = \frac{1}{d}\sum_{r=1}^{d}\delta_{\lambda_r(B)}$ with $\int f(\lambda)\,\dd\mu_B(\lambda) = \frac{1}{d}{\rm Tr}\,f(B)$. Random normalised probe vectors $z_\ell$ satisfy $\mathbb E\,|z_\ell\rangle\langle z_\ell|=\mathbb I/d$ and therefore give stochastic estimates of the normalised trace \cite{Hutchinson:1990}.  For each probe, $m_{\rm L}$ Lanczos steps applied to $B$ produce a tridiagonal matrix $T_{m_{\rm L}}^{(\ell)}$.  We denote its eigenpairs by $(\theta_s^{(\ell)},u_s^{(\ell)})$, where $s=1,\ldots,m_{\rm L}$ labels the Ritz nodes generated from the $\ell$\textsuperscript{th} probe. Note that $\theta_s^{(\ell)}$ are not asserted to be the full eigenvalue list of $B$, but rather they are quadrature nodes for the probe spectral measure. Lanczos Gaussian quadrature gives \cite{GolubMeurant:1994,UbaruChenSaad:2017}
\begin{equation}
        \langle z_\ell,f(B)z_\ell\rangle
        \simeq
        \sum_{s=1}^{m_{\rm L}}
        \abs{(u_s^{(\ell)})_1}^2 f(\theta_s^{(\ell)}).
        \label{eq:slq-gauss-quadrature}
\end{equation}
Averaging over $R$ probes gives a random quadrature measure
\begin{equation}
        \widehat\mu_{B}^{(R,m_{\rm L})}
        =
        \frac{1}{R}\sum_{\ell=1}^{R}
        \sum_{s=1}^{m_{\rm L}}
        \abs{(u_s^{(\ell)})_1}^2\,
        \delta_{\theta_s^{(\ell)}} .
        \label{eq:slq-measure}
\end{equation}

In the present calculation the operator $B$ is taken to be the finite SRQ volume action $B_M = V_{v,M}^{\rm SRQ}$. Thus, every Lanczos matrix-vector product is a direct application of the SRQ volume operator. Therefore for each given Lanczos vector $z$, one forms $\bar A_v z$ by two applications of the recoupling action of $Q_v$, and for each logarithmic shift solves the shifted system $\left(\bar A_v+e^{kh}\mathbb I\right)y_k = \bar A_v z$ and then, the weighted sum of the solutions $y_k$ is precisely $B_M z$. Consequently the tridiagonal matrices in \eqref{eq:slq-gauss-quadrature} are built from the volume action itself and no dense matrix representation of $Q_v$, $Q_v^2$, the shifted systems, or $V_v$ is used in the stochastic curves.

We then plot curves which are smoothed versions of \eqref{eq:slq-measure} for the fixed local sectors below. More explicitly, for a Gaussian kernel $G_\eta$ of bandwidth $\eta$, the displayed curve is
\begin{equation}
        \widehat\rho_{\eta}(\lambda)
        =
        \frac{1}{R}\sum_{\ell=1}^{R}
        \sum_{s=1}^{m_{\rm L}}
        \abs{(u_s^{(\ell)})_1}^2\,
        G_\eta\!\left(\lambda-\theta_s^{(\ell)}\right).
        \label{eq:plotted-slq-density}
\end{equation}
The vertical axis is therefore a normalised probability density, its integral is one and it is not an eigenvalue-count histogram over many spin sectors and sign configurations. That is, peaks of the continuous curves should not be read as individual eigenvalues but as features of a kernel-smoothed stochastic quadrature approximation to the spectral measure of $B_M$. For the two fixed sectors shown here, we also show exact histograms which are obtained separately by dense diagonalisation of the same finite local sector.

\begin{figure}[ht]
\centering
\includegraphics[width=0.62\textwidth]{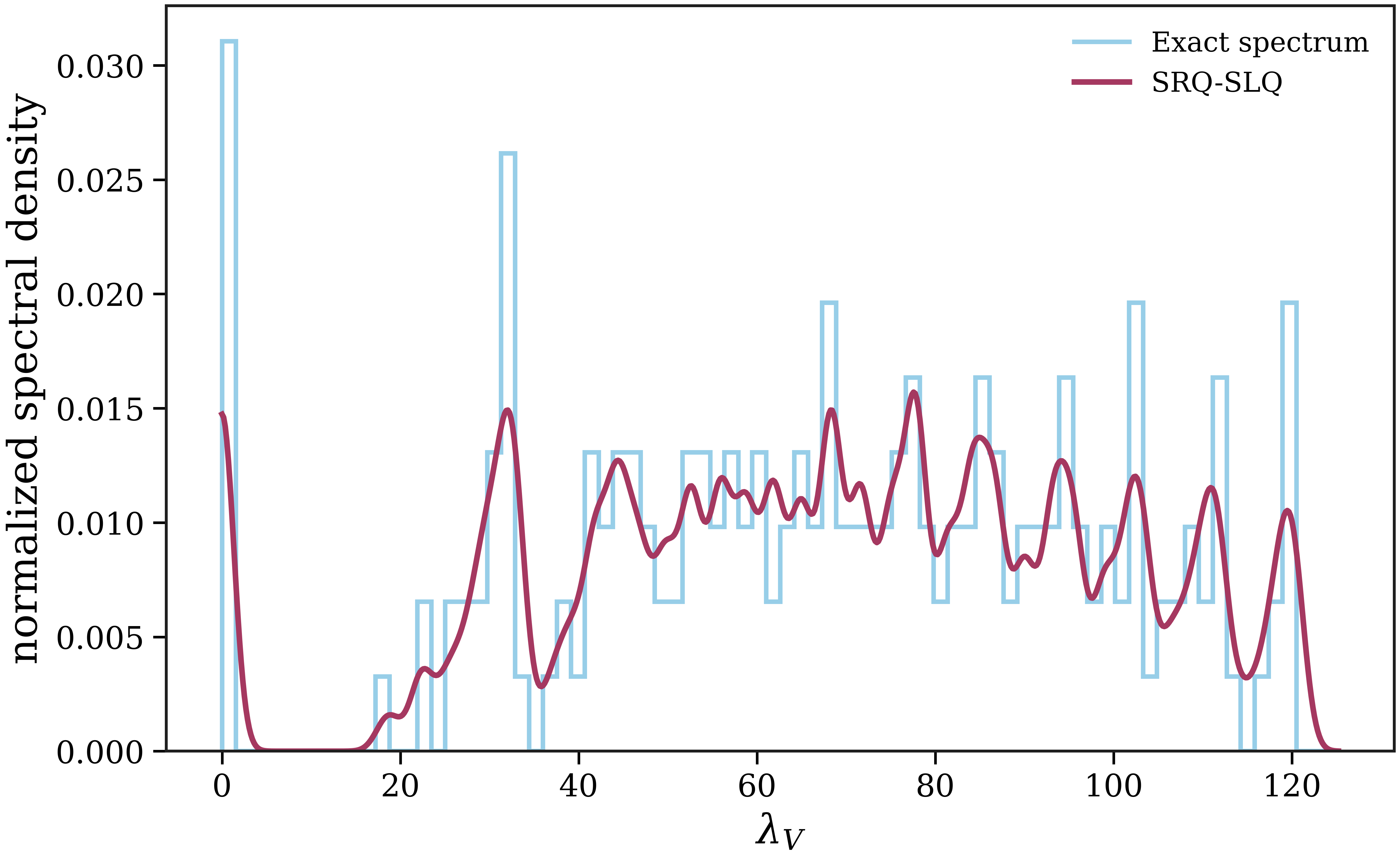}
\caption{\label{fig:five-valent-direct-srq-spin12}
Direct SRQ-SLQ spectral-density estimate for a homogeneous 5-valent spin-$12$ sector. The stochastic curve is obtained by applying SLQ directly to the finite SRQ volume action with $M=20$, $m_{\rm L}=48$, and $R=64$. The blue histogram is the exact binned finite-SRQ spectrum, included only as a reference because this sector is still small enough to diagonalise. The continuous density is a smoothed weighted Ritz measure.}
\end{figure}

\Cref{fig:five-valent-direct-srq-spin12} gives the larger of the two 5-valent tests. We take a homogeneous spin-$12$ sector, $2j_i=24$, with intertwiner dimension $391$. The coalesced recoupling action contains $2904$ sparse nonzero entries, and the row-sum bound is $\Lambda_v=1.92218847199\times10^4$. The orientation signs $\epsilon_{IJK}$ entering the Ashtekar-Lewandowski density are fixed once and used in both 5-valent spectral figures.\footnote{The 5-valent spectral tests use a realizable non-coplanar local orientation pattern generated from tangent vectors whereby in the local one-indexed leg order the signs are $\epsilon_{123}=+1$, $\epsilon_{124}=-1$, $\epsilon_{125}=+1$, $\epsilon_{134}=-1$, $\epsilon_{135}=-1$, $\epsilon_{145}=+1$, $\epsilon_{234}=+1$, $\epsilon_{235}=+1$, $\epsilon_{245}=-1$, and $\epsilon_{345}=+1$.} Here we use the intentionally modest quadrature order $M=20$, Lanczos depth $m_{\rm L}=48$, and $R=64$ probes. Even with such modest choices, the SRQ-SLQ curve places most of its spectral weight over roughly the same $\lambda_V$ range as the exact binned finite-SRQ spectrum and reproduces the broad support of the distribution. The agreement is not a bin-by-bin reconstruction, however, as some regions that carry comparatively larger exact histogram weight are under- or over-represented by the smoothed weighted curve, as expected for a finite-probe stochastic quadrature estimate.

While the preceding figure demonstrates the full direct SRQ-SLQ pipeline in a sector large enough that diagonalisation is no longer the natural computational primitive, while still small enough to permit an exact histogram for reference, it also shows that such SLQ methods are not always directly representitive of the underlying binning. To make this more explicit, we next isolate the role of the number of probes $R$. While increasing $R$ does not move the eigenvalues of $B_M$, it reduces the random error in the spectral weights assigned by the random probe ensemble.

\begin{figure}[ht]
\centering
\includegraphics[width=0.62\textwidth]{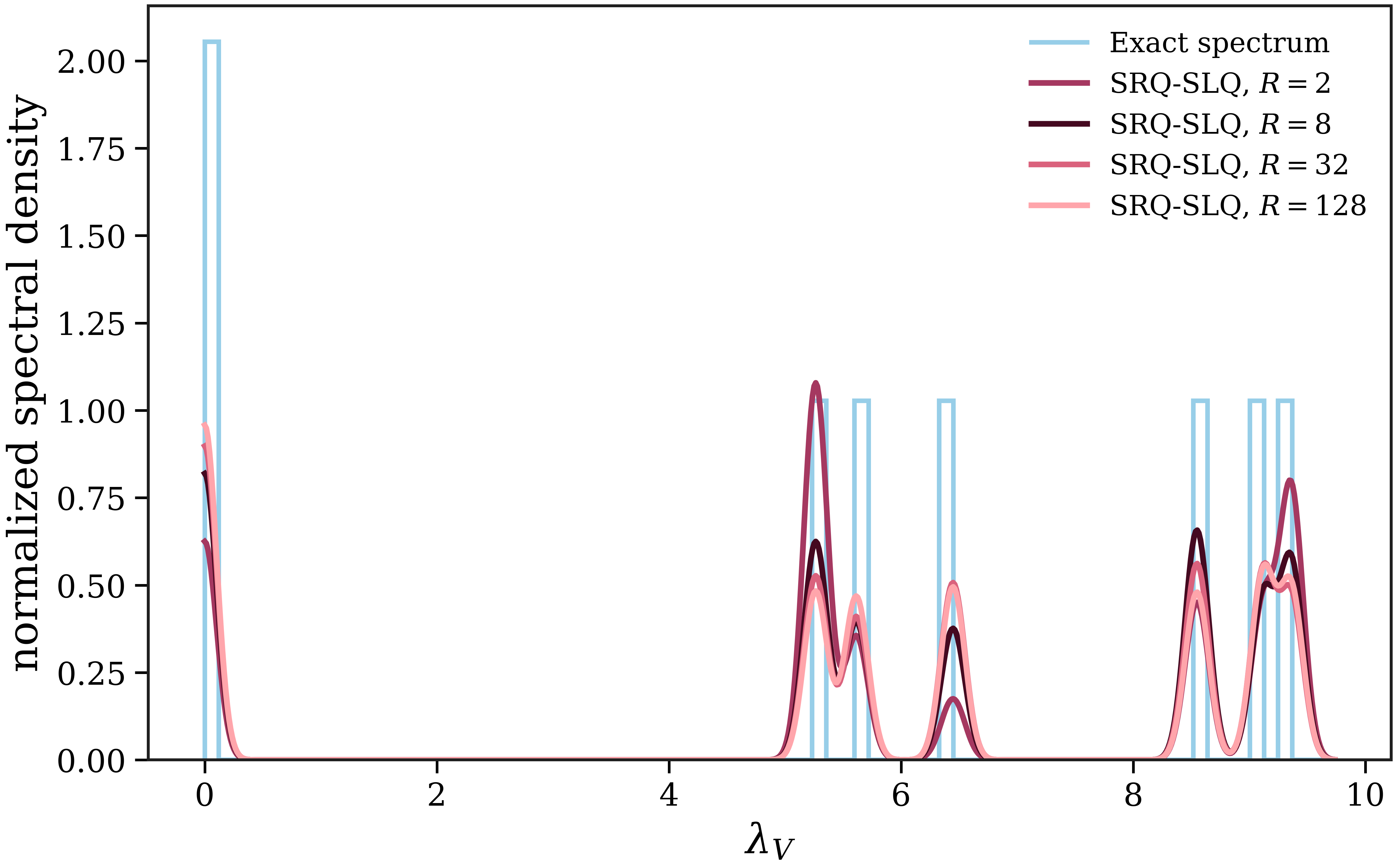}
\caption{\label{fig:five-valent-direct-srq-probes}
Probe dependence of the direct SRQ-SLQ spectral-density estimate for a homogeneous 5-valent spin-$2$ sector. The continuous curves are Gaussian-smoothed SLQ measures for the finite SRQ volume action $B_M=V_{v,M}^{\rm SRQ}$ at $M=200$, with $R=2,8,32,128$ random probes. The blue histogram is the exact binned spectrum of the same finite SRQ operator in this small sector. The comparison is at the level of the normalised spectral measure, not at the level of individual eigenvalue labels.}
\end{figure}

\Cref{fig:five-valent-direct-srq-probes} shows this effect now in the homogeneous 5-valent spin-$2$ sector, $2j_i=4$. This sector has dimension $16$, and the coalesced recoupling action contains $78$ sparse nonzero entries. The SRQ order is $M=200$, the Lanczos depth is $m_{\rm L}=16$, and the Schur row-sum bound is $\Lambda_v=137.443252114$. Since the Lanczos depth reaches the local dimension, the single-probe quadrature resolves the probe spectral measure up to roundoff and the remaining visible variation is dominated by the stochastic trace sampling over $R$. As shown in \Cref{fig:five-valent-direct-srq-probes}, changing $R$ changes the quality of the stochastic trace estimate. In particular, increasing $R$ makes the smoothed SRQ-SLQ density follow the exact finite-SRQ reference more closely, both in the location of the dominant spectral weight and in the relative weight assigned to the visible features of the distribution.

These spectral tests are therefore complementary to the expectation-value and kernel-preservation tests above. Namely, they show that the SRQ construction supplies a self-adjoint positive volume action which can be inserted directly into Krylov spectral estimators with the output being a controlled approximation to fixed-sector spectral measures and trace observables.

\section{\label{sec:resolventfree}Resolvent-free approach}

We note that the shifted resolvent construction used in this work is not the only possible integral transform representation of the non-polynomial function $\sqrt{|Q_v|}$. Indeed, a related resolvent free viewpoint, in the context of non-perturbative quantum gravity in Fock representations~\cite{ThiemannFock2024}, is directly applicable here. In~\cite{ThiemannFock2024}, the basic problem is different from the finite-dimensional recoupling problem considered here. Namely, one works with continuum metric variables in a Fock representation and has to define non-polynomial functions of the metric, including metric-density powers entering geometric observables and constraints, as quadratic forms on dense form domains. The method uses Weyl quantisation based on Fourier transforms (and normal ordering) in order to define the relevant non-polynomial quantities without passing through resolvents.

Although the representation-theoretic setting is different, the same functional calculus idea can, \emph{in spirit}, be applied to the fixed vertex problem \emph{after} the Ashtekar-Lewandowski density $Q_v$ has already been quantized as a finite-dimensional self-adjoint operator. The scalar identity
\begin{equation}
    |x|^\alpha
        =
    c_\alpha
    \int_0^\infty
    \frac{1-\cos(tx)}{t^{1+\alpha}}\,dt,
    \qquad
    0<\alpha<2,
\end{equation}
with $c_\alpha = 2\Gamma(1+\alpha)\sin(\pi\alpha/2)/\pi$, is a standard Fourier representation of fractional powers~\cite{Kwasnicki2017}.
For the volume function $\sqrt{|x|}=|x|^{1/2}$, this gives
\begin{equation}
        \sqrt{|x|}
        =
        \frac{1}{\sqrt{2\pi}}
        \int_0^\infty
        \frac{1-\cos(tx)}{t^{3/2}}\,dt .
        \label{eq:fourier-sqrt-abs-scalar}
\end{equation}
Indeed, for $x\neq0$, setting $u=t|x|$ gives
\begin{equation}
        \int_0^\infty
        \frac{1-\cos(tx)}{t^{3/2}}\,dt
        =
        |x|^{1/2}
        \int_0^\infty
        \frac{1-\cos u}{u^{3/2}}\,du
        =
        \sqrt{2\pi}\,|x|^{1/2},
\end{equation}
whereby at $x=0$, both sides vanish.

By the finite-dimensional spectral theorem, \eqref{eq:fourier-sqrt-abs-scalar} therefore yields the operator identity
\begin{equation}
        \sqrt{|Q_v|}
        =
        \frac{1}{\sqrt{2\pi}}
        \int_0^\infty
        t^{-3/2}
        \left(\mathbb{I}-\cos(tQ_v)\right)\,dt .
        \label{eq:fourier-volume-unscaled}
\end{equation}
This is an exact representation of the same Ashtekar-Lewandowski vertex
volume considered above,
\begin{equation}
        \widehat V^{\mathrm{AL}}_v
        =
        C_V\sqrt{|Q_v|}
        =
        \frac{C_V}{\sqrt{2\pi}}
        \int_0^\infty
        t^{-3/2}
        \left(\mathbb{I}-\cos(tQ_v)\right)\,dt .
        \label{eq:fourier-al-volume}
\end{equation}
The integral is harmless in the fixed vertex sector. Near $t=0$,
$\mathbb{I}-\cos(tQ_v)=O(t^2)$, so the integrand behaves as $O(t^{1/2})$. For large $t$, $|\mathbb{I}-\cos(tQ_v)\|\leq2$, so the tail is bounded by
$2t^{-3/2}$.

This representation can also be written in a positive quadratic-form form, closer \emph{in spirit} to the Weyl construction. Let $W_v(t):=e^{itQ_v}.$ Since $Q_v=Q_v^\dagger$, $W_v(t)$ is unitary and
\begin{equation}
        \mathbb{I}-\cos(tQ_v)
        =
        \frac12\left(W_v(t)-\mathbb{I}\right)^\dagger
        \left(W_v(t)-\mathbb{I}\right).
\end{equation}
Thus
\begin{equation}
        \widehat V^{\mathrm{AL}}_v
        =
        \frac{C_V}{2\sqrt{2\pi}}
        \int_0^\infty
        t^{-3/2}
        \left(e^{itQ_v}-\mathbb{I}\right)^\dagger
        \left(e^{itQ_v}-\mathbb{I}\right)\,dt .
        \label{eq:fourier-weyl-quadratic-form}
\end{equation}
This makes positivity explicit term by term.  It also makes the kernel
behaviour transparent, namely if $u\in\ker Q_v$, then
\begin{equation}
        \left(e^{itQ_v}-\mathbb{I}\right)u=0
        \qquad
        \text{for all }t,
\end{equation}
and hence every positive quadrature of \eqref{eq:fourier-weyl-quadratic-form} annihilates the exact zero-volume subspace.

For numerical comparison with SRQ, one may introduce the same sector-wise scale $\Lambda_v\geq\|Q_v\|$, assuming $\Lambda_v>0$, and set $B_v:=Q_v/\Lambda_v$ with $\sigma(B_v)\subset[-1,1]$. Then $\sqrt{|Q_v|} = \sqrt{\Lambda_v}\sqrt{|B_v|}$, and \eqref{eq:fourier-weyl-quadratic-form} becomes
\begin{equation}
        \widehat V^{\mathrm{AL}}_v
        =
        \frac{C_V\sqrt{\Lambda_v}}{2\sqrt{2\pi}}
        \int_0^\infty
        t^{-3/2}
        \left(e^{itB_v}-\mathbb{I}\right)^\dagger
        \left(e^{itB_v}-\mathbb{I}\right)\,dt ,
        \label{eq:scaled-fourier-weyl-volume}
\end{equation}
or equivalently,
\begin{equation}
        \widehat V^{\mathrm{AL}}_v
        =
        \frac{C_V\sqrt{\Lambda_v}}{\sqrt{2\pi}}
        \int_0^\infty
        t^{-3/2}
        \left(\mathbb{I}-\cos(tB_v)\right)\,dt .
        \label{eq:scaled-fourier-cos-volume}
\end{equation}
A finite resolvent-free quadrature would therefore have the schematic form
\begin{equation}
        \widehat V^{\mathrm{FW}}_{v}
        =
        \frac{C_V\sqrt{\Lambda_v}}{2\sqrt{2\pi}}
        \sum_{\ell}
        w_\ell t_\ell^{-3/2}
        \left(e^{it_\ell Q_v/\Lambda_v}-\mathbb{I}\right)^\dagger
        \left(e^{it_\ell Q_v/\Lambda_v}-\mathbb{I}\right),
        \label{eq:finite-fourier-weyl-volume}
\end{equation}
with positive quadrature weights $w_\ell$. In such a scheme the shifted
linear systems of SRQ would be replaced by matrix free applications of
$e^{itQ_v/\Lambda_v}$, for example through Lanczos, Chebyshev, or other
Krylov exponential-propagation methods. If the real antisymmetric convention $Q_v=iq_v$ is used, then $e^{itQ_v/\Lambda_v}=e^{-tq_v/\Lambda_v}$, which is a real orthogonal action because $q_v^\top=-q_v$.

At present it is not clear whether this route, even though it applies the methods of~\cite{ThiemannFock2024} in a loose manner, is also numerically robust as the shifted-resolvent method is shown to be. However, it is appealing as it avoids resolvents and replaces them by oscillatory unitary or orthogonal propagator actions and a decaying tail. By contrast, SRQ leads to positive shifted systems sharing the same Krylov space, and is therefore directly compatible with multi-shift conjugate gradient methods. A fair comparison would require a dedicated study of quadrature choices for \eqref{eq:scaled-fourier-weyl-volume}, the cost of applying $e^{itQ_v/\Lambda_v}$ in recoupling bases, and the stability of the resulting finite positive approximants, and a rigorous application of the methods introduced in~\cite{ThiemannFock2024}. We however emphasise that such a resolvent-free approach is a natural alternative backend and an interesting direction for future work, as it offers an alternative matrix free approach which does not contain multi-shift systems to be solved.

\section{\label{sec:conclusion}Conclusion}

In this work, we show that a matrix free formulation of the action of the $\mathrm{SU}(2)$ Ashtekar-Lewandowski vertex volume operator can indeed be obtained, which avoids the main obstruction in using the volume at large spin cutoff and higher valence, namely the explicit construction and diagonalisation of dense local volume matrices. The basic separation is between the algebraic recoupling action of the oriented operator $Q_v$, which can be implemented directly on spin network data, and the non-polynomial operation $\sqrt{|Q_v|}$, which is represented by a shifted-resolvent quadrature after making use of the Balakrishnan-Stieltjes representation. In this form the volume action is not introduced as a precomputed matrix, but as an operator algorithm built from repeated applications of $Q_v$ and the solution of positive shifted systems.

The numerical tests on the fixed $K_5$ graph provide a controlled validation of this construction in sectors small enough to admit an exact reference calculation. For cutoffs $j_{\max}=1$ and $j_{\max}=2$, we compared the SRQ vertex volume expectations with dense local-block volume operators evaluated by
full summation over the finite spin network basis. The presented algorithm estimates converge rapidly to the exact values as the quadrature order is increased, and the observed dependence on the sector-wise scale $\Lambda_v$ is consistent with its theoretical role as a spectral enclosure rather than a tunable physical parameter. In addition, the exact zero-volume subspace is preserved at finite quadrature order, which is an important structural property for applications in which kernel and near-kernel states carry physical significance.
The high-cutoff $K_5$ Monte Carlo test at doubled-spin cutoff $j_{\max}=250000$ then shows that the same action can be used directly in a fixed graph sector with more than $2.7\times10^{25}$ admissible basis states, where even the largest local dense $Q_v$ block would require at least $246$ GB of complex matrix storage before diagonalisation. The 5-valent spectral tests further show that this finite SRQ action can serve directly as the operator input to stochastic Lanczos quadrature, giving matrix free access to fixed-sector spectral measures and trace observables.

These results support the proposed algorithm as a faithful and scalable algorithmic definition of the vertex-volume action in regimes where dense diagonalisation is not a viable computational primitive.  The remaining, albeit straightforward, task is mainly one of high performance implementation, namely optimized recoupling kernels are needed for arbitrary valence and large spin sectors, together with efficient Krylov solvers, reliable sector-wise choices of $\Lambda_v$, and low-spectrum corrections when very small nonzero volume eigenvalues must be resolved. The framework developed here isolates these tasks while preserving the defining operator structure of the Ashtekar-Lewandowski volume, and therefore provides a practical route toward incorporating the volume operator in future variational large scale simulations.

\section{\label{sec:ack}Acknowledgments}
The author is grateful for the insightful comments and discussions with Hanno Sahlmann and Thomas Thiemann, which helped improve the presentation and sharpen several aspects of the argument.

%
%

\section*{References}
\bibliographystyle{iopart-num-long}
\bibliography{references}

@article{Rovelli:1997yv,
 author = "Rovelli, Carlo",
 title = "{Loop quantum gravity}",
 eprint = "gr-qc/9710008",
 archivePrefix = "arXiv",
 doi = "10.12942/lrr-1998-1",
 journal = "Living Rev. Rel.",
 volume = "1",
 pages = "1",
 year = "1998"
}

@book{Thiemann:2001gmi,
 author = "Thiemann, Thomas",
 title = "{Introduction to Modern Canonical Quantum General Relativity}",
 eprint = "gr-qc/0110034",
 archivePrefix = "arXiv",
 reportNumber = "AEI-2001-119, AEI-2001-119",
 year = "2001"
}

@book{Thiemann:2007pyv,
 author = "Thiemann, Thomas",
 title = "{Modern Canonical Quantum General Relativity}",
 doi = "10.1017/CBO9780511755682",
 isbn = "978-0-511-75568-2, 978-0-521-84263-1",
 publisher = "Cambridge University Press",
 series = "Cambridge Monographs on Mathematical Physics",
 year = "2007"
}

@article{Ashtekar:2004eh,
 author = "Ashtekar, Abhay and Lewandowski, Jerzy",
 title = "{Background independent quantum gravity: A Status report}",
 eprint = "gr-qc/0404018",
 archivePrefix = "arXiv",
 doi = "10.1088/0264-9381/21/15/R01",
 journal = "Class. Quant. Grav.",
 volume = "21",
 pages = "R53",
 year = "2004"
}

@article{Ashtekar:2021kfp,
    author = "Ashtekar, Abhay and Bianchi, Eugenio",
    title = "{A short review of loop quantum gravity}",
    eprint = "2104.04394",
    archivePrefix = "arXiv",
    primaryClass = "gr-qc",
    doi = "10.1088/1361-6633/abed91",
    journal = "Rept. Prog. Phys.",
    volume = "84",
    number = "4",
    pages = "042001",
    year = "2021"
}

@article{Guedes:2025taming,
  author        = {Guedes, Thiago L. M. and Mena Marug{\'a}n, Guillermo A. and M{\"u}ller, Markus and Vidotto, Francesca},
  title         = {Taming Thiemann's Hamiltonian constraint in canonical loop quantum gravity: Reversibility, eigenstates, and graph-change analysis},
  journal       = {Phys. Rev. D},
  volume        = {112},
  number        = {2},
  pages         = {026024},
  year          = {2025},
  eprint        = {2412.20272},
  archivePrefix = {arXiv},
  primaryClass  = {gr-qc},
  doi           = {10.1103/hjdk-kdhk}
}

@article{Guedes:2025graphchanging,
  author  = {Guedes, Thiago L. M. and Mena Marug{\'a}n, Guillermo A. and Vidotto, Francesca and M{\"u}ller, Markus},
  title   = {Computing the Graph-Changing Dynamics of Loop Quantum Gravity},
  journal = {Universe},
  volume  = {11},
  number  = {12},
  pages   = {387},
  year    = {2025},
  doi     = {10.3390/universe11120387}
}

@article{Sahlmann:2024pba,
    author = "Sahlmann, Hanno and Sherif, Waleed",
    title = "{Towards quantum gravity with neural networks: solving the quantum Hamilton constraint of U(1) BF theory}",
    eprint = "2402.10622",
    archivePrefix = "arXiv",
    primaryClass = "gr-qc",
    doi = "10.1088/1361-6382/ad84af",
    journal = "Class. Quant. Grav.",
    volume = "41",
    number = "22",
    pages = "225014",
    year = "2024"
}

@article{Sahlmann:2024kat,
    author = "Sahlmann, Hanno and Sherif, Waleed",
    title = "{Towards quantum gravity with neural networks: solving quantum Hamilton constraints of 3d Euclidean gravity in the weak coupling limit}",
    eprint = "2405.00661",
    archivePrefix = "arXiv",
    primaryClass = "gr-qc",
    doi = "10.1088/1361-6382/ad7c14",
    journal = "Class. Quant. Grav.",
    volume = "41",
    number = "21",
    pages = "215006",
    year = "2024"
}

@misc{Sahlmann:2026qvs,
    author = "Sahlmann, Hanno and Sherif, Waleed",
    title = "{Finding and characterising physical states of Euclidean Abelianized loop quantum gravity using neural quantum states}",
    eprint = "2604.14067",
    archivePrefix = "arXiv",
    primaryClass = "gr-qc",
    month = "4",
    year = "2026"
}

@article{Makinen:2026wwp,
    author = {M{\"a}kinen, Ilkka and Sahlmann, Hanno and Sherif, Waleed},
    title = "{Emergent Thiemann coherent states in the near-kernel sector of quantum reduced loop gravity}",
    eprint = "2605.18625",
    archivePrefix = "arXiv",
    primaryClass = "gr-qc",
    month = "5",
    year = "2026"
}

@misc{jax2018github,
 title = {{{JAX}}: Composable Transformations of {{Python}}+{{NumPy}} Programs},
 author = {Bradbury, James and Frostig, Roy and Hawkins, Peter and Johnson, Matthew James and Leary, Chris and Maclaurin, Dougal and Necula, George and Paszke, Adam and VanderPlas, Jake and Wanderman-Milne, Skye and Zhang, Qiao},
 year = {2018}
}

@article{Carleo:2016svm,
    author = "Carleo, Giuseppe and Troyer, Matthias",
    title = "{Solving the quantum many-body problem with artificial neural networks}",
    eprint = "1606.02318",
    archivePrefix = "arXiv",
    primaryClass = "cond-mat.dis-nn",
    doi = "10.1126/science.aag2302",
    journal = "Science",
    volume = "355",
    number = "6325",
    pages = "602--606",
    year = "2017"
}

@article{Ashtekar:1996eg,
 author = "Ashtekar, Abhay and Lewandowski, Jerzy",
 title = "{Quantum theory of geometry. 1: Area operators}",
 eprint = "gr-qc/9602046",
 archivePrefix = "arXiv",
 reportNumber = "CGPG-96-2-4",
 doi = "10.1088/0264-9381/14/1A/006",
 journal = "Class. Quant. Grav.",
 volume = "14",
 pages = "A55--A82",
 year = "1997"
}

@article{Rovelli:1994ge,
 author = "Rovelli, Carlo and Smolin, Lee",
 title = "{Discreteness of area and volume in quantum gravity}",
 eprint = "gr-qc/9411005",
 archivePrefix = "arXiv",
 reportNumber = "CGPG-94-11-1",
 doi = "10.1016/0550-3213(95)00150-Q",
 journal = "Nucl. Phys. B",
 volume = "442",
 pages = "593--622",
 year = "1995",
 note = "[Erratum: Nucl.Phys.B 456, 753--754 (1995)]"
}

@article{Ashtekar:1997fb,
 author = "Ashtekar, Abhay and Lewandowski, Jerzy",
 title = "{Quantum theory of geometry. 2. Volume operators}",
 eprint = "gr-qc/9711031",
 archivePrefix = "arXiv",
 reportNumber = "CGPG-97-11-1",
 doi = "10.4310/ATMP.1997.v1.n2.a8",
 journal = "Adv. Theor. Math. Phys.",
 volume = "1",
 pages = "388--429",
 year = "1998"
}

@article{DePietri:1996tvo,
 author = "De Pietri, Roberto and Rovelli, Carlo",
 title = "{Geometry eigenvalues and scalar product from recoupling theory in loop quantum gravity}",
 eprint = "gr-qc/9602023",
 archivePrefix = "arXiv",
 doi = "10.1103/PhysRevD.54.2664",
 journal = "Phys. Rev. D",
 volume = "54",
 pages = "2664--2690",
 year = "1996"
}

@article{Thiemann:1996at,
 author = "Thiemann, Thomas",
 title = "{Closed formula for the matrix elements of the volume operator in canonical quantum gravity}",
 eprint = "gr-qc/9606091",
 archivePrefix = "arXiv",
 doi = "10.1063/1.532259",
 journal = "J. Math. Phys.",
 volume = "39",
 pages = "3347--3371",
 year = "1998"
}

@article{Brunnemann:2004xi,
 author = "Brunnemann, Johannes and Thiemann, Thomas",
 title = "{Simplification of the spectral analysis of the volume operator in loop quantum gravity}",
 eprint = "gr-qc/0405060",
 archivePrefix = "arXiv",
 doi = "10.1088/0264-9381/23/4/014",
 journal = "Class. Quant. Grav.",
 volume = "23",
 pages = "1289--1346",
 year = "2006"
}

@article{brunnemann-and-rideout,
 author = "Brunnemann, Johannes and Rideout, David",
 title = "{Properties of the volume operator in loop quantum gravity. I. Results}",
 eprint = "0706.0469",
 archivePrefix = "arXiv",
 primaryClass = "gr-qc",
 doi = "10.1088/0264-9381/25/6/065001",
 journal = "Class. Quant. Grav.",
 volume = "25",
 pages = "065001",
 year = "2008"
}

@article{BrunnemannRideout:2007xk,
 author = "Brunnemann, Johannes and Rideout, David",
 title = "{Properties of the volume operator in loop quantum gravity. II. Detailed presentation}",
 eprint = "0706.0382",
 archivePrefix = "arXiv",
 primaryClass = "gr-qc",
 doi = "10.1088/0264-9381/25/6/065002",
 journal = "Class. Quant. Grav.",
 volume = "25",
 pages = "065002",
 year = "2008"
}

@article{Thiemann:1996aw,
 author = "Thiemann, Thomas",
 title = "{Anomaly-free formulation of non-perturbative, four-dimensional Lorentzian quantum gravity}",
 eprint = "gr-qc/9606088",
 archivePrefix = "arXiv",
 doi = "10.1016/0370-2693(96)00532-1",
 journal = "Phys. Lett. B",
 volume = "380",
 pages = "257--264",
 year = "1996"
}

@article{Thiemann:1996av,
 author = "Thiemann, Thomas",
 title = "{Quantum spin dynamics (QSD)}",
 eprint = "gr-qc/9606089",
 archivePrefix = "arXiv",
 doi = "10.1088/0264-9381/15/4/011",
 journal = "Class. Quant. Grav.",
 volume = "15",
 pages = "839--873",
 year = "1998"
}

@article{Thiemann:1996ay,
 author = "Thiemann, Thomas",
 title = "{Quantum spin dynamics (QSD). II. The kernel of the Wheeler-Dewitt constraint operator}",
 eprint = "gr-qc/9606090",
 archivePrefix = "arXiv",
 doi = "10.1088/0264-9381/15/4/012",
 journal = "Class. Quant. Grav.",
 volume = "15",
 pages = "875--905",
 year = "1998"
}

@article{Balakrishnan:1960,
 author = "Balakrishnan, A. V.",
 title = "{Fractional powers of closed operators and the semigroups generated by them}",
 doi = "10.2140/pjm.1960.10.419",
 journal = "Pacific J. Math.",
 volume = "10",
 number = "2",
 pages = "419--437",
 year = "1960"
}

@book{Higham:2008,
 author = "Higham, Nicholas J.",
 title = "{Functions of Matrices: Theory and Computation}",
 publisher = "Society for Industrial and Applied Mathematics",
 address = "Philadelphia, PA",
 doi = "10.1137/1.9780898717778",
 isbn = "978-0-898716-46-7",
 year = "2008"
}

@article{HaleHighamTrefethen2008,
 author = "Hale, Nicholas and Higham, Nicholas J. and Trefethen, Lloyd N.",
 title = "{Computing $A^{\alpha}$, $\log(A)$, and related matrix functions by contour integrals}",
 doi = "10.1137/070700607",
 journal = "SIAM J. Numer. Anal.",
 volume = "46",
 number = "5",
 pages = "2505--2523",
 year = "2008"
}

@article{TrefethenWeideman2014,
 author = "Trefethen, Lloyd N. and Weideman, J. A. C.",
 title = "{The exponentially convergent trapezoidal rule}",
 doi = "10.1137/130932132",
 journal = "SIAM Rev.",
 volume = "56",
 number = "3",
 pages = "385--458",
 year = "2014"
}

@article{Sorella:1998,
 author = "Sorella, Sandro",
 title = "{Green function Monte Carlo with stochastic reconfiguration}",
 eprint = "cond-mat/9803107",
 archivePrefix = "arXiv",
 doi = "10.1103/PhysRevLett.80.4558",
 journal = "Phys. Rev. Lett.",
 volume = "80",
 pages = "4558--4561",
 year = "1998"
}

@article{Sorella:2001,
 author = "Sorella, Sandro",
 title = "{Generalized Lanczos algorithm for variational quantum Monte Carlo}",
 eprint = "cond-mat/0009149",
 archivePrefix = "arXiv",
 doi = "10.1103/PhysRevB.64.024512",
 journal = "Phys. Rev. B",
 volume = "64",
 pages = "024512",
 year = "2001"
}

@article{Amari:1998,
 author = "Amari, Shun-ichi",
 title = "{Natural gradient works efficiently in learning}",
 doi = "10.1162/089976698300017746",
 journal = "Neural Comput.",
 volume = "10",
 number = "2",
 pages = "251--276",
 year = "1998"
}

@article{Baydin:2018,
 author = "Baydin, Atilim Gunes and Pearlmutter, Barak A. and Radul, Alexey Andreyevich and Siskind, Jeffrey Mark",
 title = "{Automatic differentiation in machine learning: a survey}",
 eprint = "1502.05767",
 archivePrefix = "arXiv",
 primaryClass = "cs.SC",
 journal = "J. Mach. Learn. Res.",
 volume = "18",
 number = "153",
 pages = "1--43",
 year = "2018"
}

@article{HestenesStiefel:1952,
 author = "Hestenes, Magnus R. and Stiefel, Eduard",
 title = "{Methods of conjugate gradients for solving linear systems}",
 doi = "10.6028/jres.049.044",
 journal = "J. Res. Natl. Bur. Stand.",
 volume = "49",
 number = "6",
 pages = "409--436",
 year = "1952"
}

@book{Saad:2003,
 author = "Saad, Yousef",
 title = "{Iterative Methods for Sparse Linear Systems}",
 edition = "2",
 publisher = "Society for Industrial and Applied Mathematics",
 address = "Philadelphia, PA",
 doi = "10.1137/1.9780898718003",
 isbn = "978-0-898715-34-7",
 year = "2003"
}

@misc{Jegerlehner:1996,
 author = "Jegerlehner, Beat",
 title = "{Krylov space solvers for shifted linear systems}",
 eprint = "hep-lat/9612014",
 archivePrefix = "arXiv",
 year = "1996"
}

@article{FrommerGlassner:1998,
 author = "Frommer, Andreas and Gl{\"a}ssner, Uwe",
 title = "{Restarted GMRES for shifted linear systems}",
 doi = "10.1137/S1064827596304563",
 journal = "SIAM J. Sci. Comput.",
 volume = "19",
 number = "1",
 pages = "15--26",
 year = "1998"
}

@article{BaumannVanGijzen:2015,
 author = "Baumann, Manuel and van Gijzen, Martin B.",
 title = "{Nested Krylov methods for shifted linear systems}",
 doi = "10.1137/140979927",
 journal = "SIAM J. Sci. Comput.",
 volume = "37",
 number = "5",
 pages = "S90--S112",
 year = "2015"
}

@article{Lanczos:1950,
 author = "Lanczos, Cornelius",
 title = "{An iteration method for the solution of the eigenvalue problem of linear differential and integral operators}",
 doi = "10.6028/jres.045.026",
 journal = "J. Res. Natl. Bur. Stand.",
 volume = "45",
 number = "4",
 pages = "255--282",
 year = "1950"
}

@book{HornJohnson:2012,
 author = "Horn, Roger A. and Johnson, Charles R.",
 title = "{Matrix Analysis}",
 edition = "2",
 publisher = "Cambridge University Press",
 doi = "10.1017/CBO9781139020411",
 isbn = "978-0-521-83940-2",
 year = "2012"
}

@article{Knyazev:2001,
 author = "Knyazev, Andrew V.",
 title = "{Toward the optimal preconditioned eigensolver: locally optimal block preconditioned conjugate gradient method}",
 doi = "10.1137/S1064827500366124",
 journal = "SIAM J. Sci. Comput.",
 volume = "23",
 number = "2",
 pages = "517--541",
 year = "2001"
}

@book{SaadEigen:2011,
  author    = {Saad, Yousef},
  title     = {Numerical Methods for Large Eigenvalue Problems},
  edition   = {Revised},
  publisher = {Society for Industrial and Applied Mathematics},
  address   = {Philadelphia},
  year      = {2011},
  doi       = {10.1137/1.9781611970739}
}

@article{SleijpenVanderVorst:1996,
  author  = {Sleijpen, Gerard L. G. and van der Vorst, Henk A.},
  title   = {A Jacobi--Davidson Iteration Method for Linear Eigenvalue Problems},
  journal = {SIAM Journal on Matrix Analysis and Applications},
  volume  = {17},
  number  = {2},
  pages   = {401--425},
  year    = {1996},
  doi     = {10.1137/S0895479894270427}
}

@article{Guettel:2013,
  author  = {G{\"u}ttel, Stefan},
  title   = {Rational Krylov Approximation of Matrix Functions: Numerical Methods and Optimal Pole Selection},
  journal = {GAMM-Mitteilungen},
  volume  = {36},
  number  = {1},
  pages   = {8--31},
  year    = {2013},
  doi     = {10.1002/gamm.201310002}
}

@article{NakatsukasaFreund:2016,
  author  = {Nakatsukasa, Yuji and Freund, Roland W.},
  title   = {Computing Fundamental Matrix Decompositions Accurately via the Matrix Sign Function in Two Iterations: The Power of Zolotarev's Functions},
  journal = {SIAM Review},
  volume  = {58},
  number  = {3},
  pages   = {461--493},
  year    = {2016},
  doi     = {10.1137/140990334}
}

@article{Hutchinson:1990,
  author  = {Hutchinson, M. F.},
  title   = {A stochastic estimator of the trace of the influence matrix for Laplacian smoothing splines},
  journal = {Communications in Statistics - Simulation and Computation},
  volume  = {19},
  number  = {2},
  pages   = {433--450},
  year    = {1990},
  doi     = {10.1080/03610919008812866}
}

@incollection{GolubMeurant:1994,
  author    = {Golub, Gene H. and Meurant, G{\'e}rard},
  title     = {Matrices, moments and quadrature},
  booktitle = {Numerical Analysis 1993},
  editor    = {Griffiths, D. F. and Watson, G. A.},
  series    = {Pitman Research Notes in Mathematics Series},
  volume    = {303},
  pages     = {105--156},
  publisher = {Longman Scientific \& Technical},
  address   = {Harlow},
  year      = {1994}
}

@article{UbaruChenSaad:2017,
  author  = {Ubaru, Shashanka and Chen, Jie and Saad, Yousef},
  title   = {Fast estimation of $\operatorname{tr}(f(A))$ via stochastic Lanczos quadrature},
  journal = {SIAM Journal on Matrix Analysis and Applications},
  volume  = {38},
  number  = {4},
  pages   = {1075--1099},
  year    = {2017},
  doi     = {10.1137/16M1104974}
}

@article{Sahlmann:2002qj,
 author = "Sahlmann, Hanno and Thiemann, Thomas",
 title = "{Towards the QFT on curved space-time limit of QGR. 2. A concrete implementation}",
 eprint = "gr-qc/0207031",
 archivePrefix = "arXiv",
 doi = "10.1088/0264-9381/23/3/020",
 journal = "Class. Quant. Grav.",
 volume = "23",
 pages = "909--954",
 year = "2006"
}

@article{Giesel:2006uk,
 author = "Giesel, K. and Thiemann, T.",
 title = "{Algebraic Quantum Gravity (AQG). III. Semiclassical perturbation theory}",
 eprint = "gr-qc/0607101",
 archivePrefix = "arXiv",
 doi = "10.1088/0264-9381/24/10/005",
 journal = "Class. Quant. Grav.",
 volume = "24",
 pages = "2565--2588",
 year = "2007"
}

@article{ThiemannFock2024,
  author        = {Thiemann, Thomas},
  title         = {Non-perturbative Quantum Gravity in Fock representations},
  journal       = {Physical Review D},
  volume        = {110},
  number        = {12},
  pages         = {124023},
  year          = {2024},
  doi           = {10.1103/PhysRevD.110.124023},
  eprint        = {2405.01212},
  archivePrefix = {arXiv},
  primaryClass  = {gr-qc}
}

@article{Kwasnicki2017,
  author  = {Kwa{\'s}nicki, Mateusz},
  title   = {Ten equivalent definitions of the fractional Laplace operator},
  journal = {Fractional Calculus and Applied Analysis},
  volume  = {20},
  number  = {1},
  pages   = {7--51},
  year    = {2017},
  doi     = {10.1515/fca-2017-0002}
}

%
%

\appendix

\section{Convergence, error bounds and parameter selection}
\label{sec:errorbounds}

This appendix presents the standard scalar and operator estimates used to choose the SRQ parameters. The fractional power identity is the Balakrishnan-Stieltjes representation for positive operators, and the logarithmic quadrature estimates are standard consequences of the exponentially convergent trapezoidal rule for functions analytic in a strip (see, for example, \cite{Balakrishnan:1960,Higham:2008,HaleHighamTrefethen2008,TrefethenWeideman2014}). We specialize those standard estimates to the positive scaled operator $0\leq \bar A \leq 1$, and to the exponent relevant for the Ashtekar-Lewandowski volume, $\alpha=1/4$. For completeness, we keep the formulae explicit, since these are the constants entering the numerical implementation.

For a general exponent $0<\alpha<1$, set $c_\alpha:=\sin(\pi\alpha)/{\pi}$. The logarithmic form of the Stieltjes representation gives, for $0\leq\lambda\leq1$,
\begin{equation}
        \lambda^\alpha
        =
        c_\alpha\int_{-\infty}^{\infty}
        e^{\alpha s}\frac{\lambda}{e^s+\lambda}\,\dd s.
\end{equation}
Thus define
\begin{equation}
        g_\lambda(s)
        :=
        e^{\alpha s}\frac{\lambda}{e^s+\lambda},
        \qquad
        0\leq\lambda\leq1 .
\end{equation}
With grid spacing $h>0$, cutoffs $K_-,K_+\in\mathbb N_0$, and shifts
\begin{equation}
        \tau_k:=e^{kh},
        \qquad
        k=-K_-,\ldots,K_+,
\end{equation}
the corresponding finite rational approximant is
\begin{equation}
        r^{(\alpha)}_{h,K_-,K_+}(\lambda)
        :=
        c_\alpha h
        \sum_{k=-K_-}^{K_+}
        e^{\alpha kh}
        \frac{\lambda}{e^{kh}+\lambda}.
        \label{eq:scalar-srq-appendix}
\end{equation}
At the operator level this is
\begin{equation}
        R^{(\alpha)}_{h,K_-,K_+}(\bar A)
        :=
        c_\alpha h
        \sum_{k=-K_-}^{K_+}
        e^{\alpha kh}
        (\bar A+e^{kh}\mathbb I)^{-1}\bar A .
        \label{eq:operator-srq-appendix}
\end{equation}
Each inverse in \eqref{eq:operator-srq-appendix} is a shifted inverse.  It is therefore well-defined even when $\bar A$ has a kernel, since $\bar A+e^{kh}\mathbb I\geq e^{kh}\mathbb I>0$. The standard strip-analytic trapezoidal estimate applied to $g_\lambda$ yields the following uniform bound.  Define
\begin{equation}
        I_\alpha
        :=
        \int_0^\infty\frac{u^{\alpha-1}}{\sqrt{1+u^2}}\,\dd u
        =
        \frac12 B\left(\frac\alpha2,\frac{1-\alpha}{2}\right).
\end{equation}
Then, uniformly for $0\leq\lambda\leq1$,
\begin{align}
        \left|
        \lambda^\alpha
        -
        r^{(\alpha)}_{h,K_-,K_+}(\lambda)
        \right|
        &\leq
        E_{\rm disc}(h)+E_{\rm left}(h,K_-)+E_{\rm right}(h,K_+),
        \label{eq:scalar-error-bound-appendix}
\end{align}
where
\begin{align}
        E_{\rm disc}(h)
        &:={}
        c_\alpha\frac{2I_\alpha}{e^{\pi^2/h}-1},
        \label{eq:edisc-appendix}\\
        E_{\rm left}(h,K_-)
        &:={}
        c_\alpha h
        \frac{e^{-\alpha(K_-+1)h}}{1-e^{-\alpha h}},
        \label{eq:eleft-appendix}\\
        E_{\rm right}(h,K_+)
        &:={}
        c_\alpha h
        \frac{e^{-(1-\alpha)(K_++1)h}}{1-e^{-(1-\alpha)h}}.
        \label{eq:eright-appendix}
\end{align}
The three terms respectively correspond to the infinite-grid discretization error and the two tails of the logarithmic grid.  The constant $E_{\rm disc}$ is obtained by using the horizontal strip $|\operatorname{Im}z|<\pi$ and evaluating the usual boundary $L^1$ estimate on the smaller strip $|\operatorname{Im}z|=\pi/2$, the tail estimates follow from the elementary inequalities
\begin{equation}
        g_\lambda(s)\leq e^{\alpha s}\quad(s<0),
        \qquad
        g_\lambda(s)\leq e^{-(1-\alpha)s}\quad(s>0),
\end{equation}
which are simply the standard sinc/logarithmic quadrature estimates specialized to the Stieltjes integrand.

Since \eqref{eq:scalar-error-bound-appendix} is uniform in the spectral parameter and $\bar A$ is self-adjoint with spectrum contained in $[0,1]$, the finite-dimensional spectral theorem gives immediately
\begin{equation}
        \norm{\bar A^\alpha-R^{(\alpha)}_{h,K_-,K_+}(\bar A)}
        \leq
        E_{\rm disc}(h)+E_{\rm left}(h,K_-)+E_{\rm right}(h,K_+).
        \label{eq:operator-error-bound-general}
\end{equation}
Thus $R^{(\alpha)}_{h,K_-,K_+}(\bar A)$ converges to $\bar A^\alpha$ in operator norm as $h\to0$, $K_-h\to\infty$, $K_+h\to\infty$.

\subsection{Residual error from inexact shifted solves}
\label{subsec:residualerror}

In the SRQ implementation, the exact shifted vectors are $y_k=(\bar A_v+\tau_k\mathbb I)^{-1}g$ with $g=\bar A_v\psi$ and $\tau_k=e^{kh}$. Suppose that the iterative solver returns approximate solutions $\widetilde y_k$ with residuals $r_k:=g-(\bar A_v+\tau_k\mathbb I)\widetilde y_k$. Then $y_k-\widetilde y_k=(\bar A_v+\tau_k\mathbb I)^{-1}r_k$ and the positivity of $\bar A_v$ implies $\norm{(\bar A_v+\tau_k\mathbb I)^{-1}} \leq \tau_k^{-1}$. Consequently the shifted-solve contribution to the scaled approximation error obeys the a posteriori estimate
\begin{equation}
        \left\|
        c_\alpha h\sum_{k=-K_-}^{K_+}e^{\alpha kh}(y_k-\widetilde y_k)
        \right\|
        \leq
        c_\alpha h\sum_{k=-K_-}^{K_+}e^{\alpha kh}\frac{\norm{r_k}}{\tau_k}.
        \label{eq:solve-error-bound}
\end{equation}
For the Ashtekar-Lewandowski volume, $\alpha=1/4$, and therefore
\begin{equation}
        c_{1/4} h\sum_{k=-K_-}^{K_+}e^{kh/4}\frac{\norm{r_k}}{e^{kh}}
        =
        c_{1/4}h\sum_{k=-K_-}^{K_+}e^{-3kh/4}\norm{r_k}.
\end{equation}
This residual estimate is conservative for small shifts. In practice, one may combine it with standard conjugate gradient energy norm estimates and adaptive tolerances per shift (see, for example, \cite{HestenesStiefel:1952,Saad:2003}). Low-spectrum deflation, discussed below, is another way of avoiding oversolving shifts whose contribution is dominated by already-resolved spectral data.

\subsection{Parameter selection}
\label{subsec:parameterselection}

Let $M := K_- + K_+ + 1$ be the number of shifted systems. For $\alpha=1/4$, the left and right tails in \eqref{eq:eleft-appendix}-\eqref{eq:eright-appendix} decay respectively like $e^{-K_-h/4}$ and $e^{-3K_+h/4}$. Balancing these two tail exponents gives
\begin{equation}
        K_-\approx\frac34(M-1),
        \qquad
        K_+\approx\frac14(M-1),
        \label{eq:K-choice}
\end{equation}
and balancing the common tail exponent $3Mh/16$ against the strip-discretization exponent $\pi^2/h$ gives the practical choice
\begin{equation}
        h\approx\frac{4\pi}{\sqrt{3M}}.
        \label{eq:h-choice}
\end{equation}
With this choice the standard logarithmic-quadrature estimate gives root-exponential behaviour, up to algebraic factors,
\begin{equation}
        \norm{\bar A^{1/4}-R^{(1/4)}_M(\bar A)}
        \lesssim
        \exp\left(-\frac{\pi\sqrt3}{4}\sqrt M\right).
        \label{eq:root-exponential}
\end{equation}
Here the symbol $\lesssim$ indicates the asymptotic scale obtained by balancing the exponents. The rigorous bound used for error control remains \eqref{eq:operator-error-bound-general} with the explicit terms \eqref{eq:edisc-appendix}-\eqref{eq:eright-appendix}.

Returning to the unscaled Ashtekar-Lewandowski vertex volume, let $\rho_v:=\norm{Q_v}$. If $\rho_v>0$, then $\norm{\widehat V_v^{\rm AL}}=C_V\sqrt{\rho_v}$, and the scaled error bound implies
\begin{equation}
        \frac{
        \norm{\widehat V_v^{\rm AL}-\widehat V^{\rm SRQ}_{v,h,K_-,K_+}}
        }{
        \norm{\widehat V_v^{\rm AL}}
        }
        \leq
        \sqrt{\frac{\Lambda_v}{\rho_v}}
        \left(E_{\rm disc}+E_{\rm left}+E_{\rm right}\right).
        \label{eq:relative-error-Lambda}
\end{equation}
A tight sector scale $\Lambda_v$ is therefore still useful.  However, unlike raw Chebyshev approximants to $\sqrt{|x|}$, a conservative $\Lambda_v$ does not produce a finite zero-mode lift, because every SRQ summand contains the factor $\bar A_v$.

\section{Choosing a scaling $\Lambda_v$}
\label{sec:choosinglambda}

For the Ashtekar-Lewandowski volume, the SRQ construction requires a sector-wise number $\Lambda_v(\bm j,T)>0$ satisfying
\begin{equation}
        \rho_v(\bm j,T)
        :=
        \norm{Q_v}_{\Hcal_v(\bm j)\to\Hcal_v(\bm j)}
        \leq
        \Lambda_v(\bm j,T).
\end{equation}
This is a local condition.  The operator $Q_v$ preserves the incident spins $\bm j=(j_1,\ldots,j_N)$ and acts only on the finite-dimensional intertwiner space $\Hcal_v(\bm j)$.  Thus one should normally use a family of local sector bounds rather than a single graph-wide bound.  A global bound is obtained by maximizing over sectors, but this is usually unnecessarily loose. We collect several practical choices. The estimates use only standard matrix norm inequalities \cite{HornJohnson:2012} and their role here is to provide safe sector scales for the SRQ normalization.

Recall the gauge-reduced form
\begin{equation}
        Q_v
        =
        \sum_{1\leq I<J<K<N}\sigma_{IJK}(v)G_{IJK},
        \qquad
        G_{IJK}=\epsilon_{ijk}X_I^iX_J^jX_K^k .
\end{equation}
Let $s_I:=\sqrt{j_I(j_I+1)}$. Since operators on different tensor factors commute and $\norm{X_I^i}\leq s_I$, the elementary componentwise estimate gives $\norm{G_{IJK}} \leq 6s_Is_Js_K$. Consequently
\begin{equation}
        \Lambda^{\rm prod}_v(\bm j)
        :=
        6\sum_{I<J<K<N}
        |\sigma_{IJK}(v)|
        \sqrt{j_I(j_I+1)j_J(j_J+1)j_K(j_K+1)}
        \label{eq:lambda-product-bound}
\end{equation}
is a valid a priori upper bound for $\rho_v(\bm j,T)$ in the normalization of $Q_v$ used above.  This estimate is independent of the recoupling tree and requires no spectral information.  It also retains the AL embedding data through the reduced orientation coefficients $\sigma_{IJK}(v)$, so triples with vanishing reduced orientation coefficient do not increase the bound.  If the eliminated edge in the gauge reduction is not fixed by convention, the same estimate may be evaluated for each allowed eliminated edge and the minimum of the resulting upper bounds may be used.

A second a priori estimate is useful when the triple action is implemented through commutators of pair Casimirs, as in the standard recoupling formulas for the volume-density matrix elements \cite{DePietri:1996tvo,Brunnemann:2004xi,brunnemann-and-rideout}.  For example, for
\begin{equation}
        q^{\rm comm}_{IJK}
        =
        \big[(X_I+X_J)^2,(X_J+X_K)^2\big],
\end{equation}
then $\norm{[B,C]} \leq 2\norm{B}\,\norm{C}$, and $\norm{(X_I+X_J)^2} \leq (j_I+j_J)(j_I+j_J+1)$. Thus, after inserting the convention-dependent factor relating the commutator form to $G_{IJK}$, one obtains the safe estimate
\begin{equation}
        \Lambda^{\rm comm}_v(\bm j)
        :=
        2\sum_{I<J<K<N}
        |\sigma_{IJK}(v)|
        C_{IJ}(\bm j)C_{JK}(\bm j),
        \qquad
        C_{IJ}(\bm j):=(j_I+j_J)(j_I+j_J+1).
        \label{eq:lambda-comm-bound}
\end{equation}
This bound is typically conservative, but it is simple and remains valid at arbitrary valence once the same normalization of $Q_v$ is used throughout.

The a priori estimates above are safe but may be loose.  A sharper bound can be obtained by a streaming Schur estimate without materialising the dense matrix. Let $\{\ket{\bm a}\}_{\bm a\in\Acal_{\bm j,T}}$ be the chosen orthonormal recoupling basis, and let $Q_{\bm b\bm a}:=\langle\bm b, Q_v\,\bm a\rangle_T$. Define the absolute column and row sums
\begin{equation}
        c_{\bm a}:=
        \sum_{\bm b\in\Acal_{\bm j,T}}|Q_{\bm b\bm a}|,
        \qquad
        r_{\bm b}:=
        \sum_{\bm a\in\Acal_{\bm j,T}}|Q_{\bm b\bm a}|.
\end{equation}
These sums can be accumulated by traversing the finite local action graph generated by the recoupling formulas.  For each input channel $\bm a$, only the admissible output channels generated by the local triple graspings contribute.  The full dense matrix need not be stored.

The Schur norm bound gives
\begin{equation}
        \rho_v(\bm j,T)=\norm{Q_v}_2
        \leq
        \sqrt{\norm{Q_v}_1\norm{Q_v}_\infty}
        =
        \sqrt{
        \left(\max_{\bm a}c_{\bm a}\right)
        \left(\max_{\bm b}r_{\bm b}\right)
        }
        =:
        \Lambda^{\rm Schur}_v(\bm j,T).
        \label{eq:lambda-schur-bound}
\end{equation}
Because $Q_v$ is self-adjoint, the absolute row and column norms coincide in exact arithmetic, and the last expression may be replaced by the maximum absolute row sum.  This bound is often substantially tighter than \eqref{eq:lambda-product-bound} or \eqref{eq:lambda-comm-bound}, while its memory cost is linear in the number of channels.  Its computational cost is proportional to the number of locally generated nonzero matrix elements, not to $d_v(\bm j)^2$.

A matrix free Krylov calculation can also be used to estimate the scale.  Lanczos applied to $Q_v$, or to the positive operator $Q_v^2$, produces Ritz values approximating the extremal spectrum \cite{Lanczos:1950,Saad:2003}.  Such Ritz data are valuable as diagnostics of the looseness of an analytic or Schur bound.  However, an unconverged Krylov estimate is not, by itself, a certified upper bound for $\rho_v$.  It should therefore be used either only diagnostically, or together with an a posteriori residual enclosure or an independent safe bound. In numerical work one may report $\chi_v:=\Lambda_v/\widehat\rho_v^{\rm Ritz}$ as a tightness diagnostic, where $\widehat\rho_v^{\rm Ritz}$ is the largest available Ritz estimate below the true spectral radius.

A robust sector-wise prescription is
\begin{equation}
        \Lambda_v(\bm j,T)
        =
        (1+\varepsilon_{\rm saf})
        \min\left\{
        \Lambda^{\rm prod}_v(\bm j),
        \Lambda^{\rm comm}_v(\bm j),
        \Lambda^{\rm Schur}_v(\bm j,T)
        \right\},
        \label{eq:lambda-recommended}
\end{equation}
where only bounds that have actually been evaluated are included, all bounds are expressed in the same normalization of $Q_v$, and $\varepsilon_{\rm saf}>0$ is a small floating-point safety margin.  In exact arithmetic one may set $\varepsilon_{\rm saf}=0$.  Since the minimum of certified upper bounds is again a certified upper bound, \eqref{eq:lambda-recommended} remains safe.

If all evaluated upper bounds vanish, then every contributing orientation or grasping term vanishes on the sector and $Q_v=0$.  The vertex volume is then exactly zero on that sector and no shifted systems are required.  Otherwise, any positive $\Lambda_v\geq\rho_v$ gives a valid SRQ operator.  A loose bound enlarges the interval on which the scalar approximation is used and appears in \eqref{eq:relative-error-Lambda} through the factor $\sqrt{\Lambda_v/\rho_v}$.  It does not change the exact annihilation of $\ker Q_v$, because every SRQ summand still contains the factor $\bar A_v$.

\section{Low-spectrum correction and near-kernel states}
\label{sec:nearkernelstates}

The finite SRQ approximant protects the exact kernel, but near-kernel accuracy is a separate issue.  The function $\lambda^{1/4}$ has a branch point at $\lambda=0$, and no finite analytic approximant can have uniformly small relative error for all $0<\lambda\ll1$ without resolving the small spectral scale.  In physical and variational applications, near-zero volume sectors may be important. A natural strategy is therefore a low-rank spectral correction combined with a rational or SRQ approximation on the complementary gapped interval.  This is standard numerical linear algebra practice for matrix functions and extremal spectral components \cite{Higham:2008,SaadEigen:2011,Guettel:2013}.

Let $P_{\rm low}$ be the spectral projector of $\bar A_v$ onto $[0,\delta]$, with $0<\delta<1$.  Then
\begin{equation}
        \bar A_v^{1/4}\psi
        =
        \bar A_v^{1/4}P_{\rm low}\psi
        +
        \bar A_v^{1/4}(\mathbb I-P_{\rm low})\psi.
        \label{eq:spectral-split}
\end{equation}
On $(\mathbb I-P_{\rm low})\Hcal_v$, the spectrum is contained in $[\delta,1]$, where $\lambda^{1/4}$ is analytic in a neighbourhood of the interval and rational or polynomial approximation is considerably easier.

In practice, the low subspace may be computed by a matrix free eigensolver applied to $\bar A_v$, such as Lanczos, locally optimal block preconditioned conjugate gradient (LOBPCG), or Jacobi-Davidson methods \cite{Lanczos:1950,Knyazev:2001,SleijpenVanderVorst:1996,SaadEigen:2011}.  This is not full diagonalization of the vertex block, rather it is a low-rank correction using only the action of $\ApplyA_v$.

If the computed low eigenpairs are $\bar A_vu_r=\mu_ru_r$, $0\leq\mu_r\leq\delta$ with $r=1,\ldots,s$ and orthonormal $u_r$, then the hybrid action is
\begin{align}
        \widehat V^{\rm hybrid}_{v}\psi
        :=
        C_V\sqrt{\Lambda_v}
        \left[
        \sum_{r=1}^s\mu_r^{1/4}u_r\langle u_r,\psi\rangle
        +
        R_{\rm gap}\!
        \left((\mathbb I-P_{\rm low})\bar A_v(\mathbb I-P_{\rm low})\right)
        (\mathbb I-P_{\rm low})\psi
        \right],
        \label{eq:lowspec-corrected}
\end{align}
where $P_{\rm low}=\sum_{r=1}^s|u_r\rangle\langle u_r|$ and $R_{\rm gap}$ is a rational, SRQ, Zolotarev-type, or polynomial approximation to $\lambda^{1/4}$ on $[\delta,1]$ \cite{Higham:2008,HaleHighamTrefethen2008,Guettel:2013,NakatsukasaFreund:2016}.  Zero eigenvalues among the $\mu_r$ are assigned exactly zero.  For high-valent LQG vertices, this hybrid strategy is likely the most robust implementation as it treats the small low spectrum explicitly, applies a matrix free rational approximation to the bulk, and preserves the exact kernel.

\subsection{Finite-iteration kernel preservation}
\label{subsec:finiteiterationkernel}

The exact SRQ summand $(\bar A_v+\tau\mathbb I)^{-1}\bar A_v$ annihilates $\ker\bar A_v$ because the rightmost factor is $\bar A_v$.  The same protection is retained by zero-start Krylov solves in exact arithmetic.  Indeed, let $g=\bar A_v\psi$. Since $\bar A_v$ is self-adjoint and finite-dimensional, ${\rm Ran}\,\bar A_v=(\ker\bar A_v)^\perp$. Hence $g\in(\ker\bar A_v)^\perp$, and this subspace is invariant under $\bar A_v$ and under $\bar A_v+\tau\mathbb I$.  A zero-start Krylov iterate for $(\bar A_v+\tau\mathbb I)y=g$ belongs to
\begin{equation}
        \mathcal K_n(\bar A_v,g)
        =
        \operatorname{span}\{g,\bar A_vg,\ldots,\bar A_v^{n-1}g\}
        \subset
        (\ker\bar A_v)^\perp.
\end{equation}
If $\psi\in\ker\bar A_v$, then $g=0$ and every Krylov vector is zero.  Thus finite solver iteration does not create a component in the exact kernel, provided exact arithmetic, zero initial guesses, and kernel-preserving preconditioning are used (see \cite{HestenesStiefel:1952,Saad:2003} for standard conjugate gradient and Krylov methods).

In floating point arithmetic, preconditioners should be chosen so as not to inject components in $\ker\bar A_v$.  A safe alternative, when kernel contamination is suspected, is the algebraically equivalent protected form
\begin{equation}
        \bar A_v(\bar A_v+\tau\mathbb I)^{-1}\psi,
        \label{eq:sandwich-safe}
\end{equation}
or a final application of $\bar A_v$.  Since $\bar A_v$ commutes with its resolvent, \eqref{eq:sandwich-safe} is identical to the original exact summand, but the final factor explicitly kills any numerical component that has leaked into the kernel.

\end{document}